\documentclass[]{aastex63}
\usepackage{}
\usepackage{multirow}
\usepackage{graphicx}
\usepackage{amsmath,amssymb}
\usepackage{longtable}

\graphicspath{{./}{figures/}}

\begin{document}

\title{Photometric and Spectroscopic Observations of GRB~210104A: Bright Reverse shock Emission and Dense Circumburst Environment}
\correspondingauthor{Li-Ping Xin; En-Wei Liang}
\email{xlp@nao.cas.cn; lew@gxu.edu.cn}
\author[0000-0003-0726-7579]{Lu-Lu Zhang}
\affil{Guangxi Key Laboratory for Relativistic Astrophysics, School of Physical Science and Technology, Guangxi University, Nanning 530004, China}
\author[0000-0002-6162-9115]{Li-Ping Xin*}
\affil{Key Laboratory of Space Astronomy and Technology, National Astronomical Observatories, Chinese Academy of Sciences, Beijing 100101, China}
\author[0000-0002-6880-4481]{Jing Wang}
\affil{Guangxi Key Laboratory for Relativistic Astrophysics, School of Physical Science and Technology, Guangxi University, Nanning 530004, China}
\affil{Key Laboratory of Space Astronomy and Technology, National Astronomical Observatories, Chinese Academy of Sciences, Beijing 100101, China}
\author{Xu-Hui Han}
\affil{Key Laboratory of Space Astronomy and Technology, National Astronomical Observatories, Chinese Academy of Sciences, Beijing 100101, China}
\author{Dong Xu}
\affil{Key Laboratory of Space Astronomy and Technology, National Astronomical Observatories, Chinese Academy of Sciences, Beijing 100101, China}
\author{Zi-Pei Zhu}
\affil{Key Laboratory of Space Astronomy and Technology, National Astronomical Observatories, Chinese Academy of Sciences, Beijing 100101, China}
\author{Chao Wu}
\affil{Key Laboratory of Space Astronomy and Technology, National Astronomical Observatories, Chinese Academy of Sciences, Beijing 100101, China}
\author{Jian-Yan Wei}
\affil{Key Laboratory of Space Astronomy and Technology, National Astronomical Observatories, Chinese Academy of Sciences, Beijing 100101, China}
\author[0000-0002-7044-733X]{En-Wei Liang*}
\affil{Guangxi Key Laboratory for Relativistic Astrophysics, School of Physical Science and Technology, Guangxi University, Nanning 530004, China}

\begin{abstract}
Early afterglow observations of gamma-ray bursts (GRBs) are valuable for exploring the properties of their jets and ambient medium.
We report our photometric and spectroscopic observations of GRB 210104A and discuss its jet properties with multiwavelength data.
Our spectroscopic observation reveals several absorption features and a tentative redshift of 0.46 is identified.
A bright optical flare that has a peak brightness of $R=13$ mag at $112\pm 7$~s was observed in the $R$ band during $67\sim 165$ seconds post the GRB trigger.
The flux of the $R$-band afterglow decays with a slope of $\alpha_{\rm O}={-0.91\pm 0.03}$ at $t>650$~s.
The early X-ray afterglow lightcurve is a smooth bump, and it decays with a slope of $\alpha_{\rm X}=-1.18\pm 0.01$ at late epoch.
Our joint spectral fit to the optical-X-ray afterglows during $(1.1-1.3)\times 10^{4}$~s yields a photon index $\Gamma_{\rm O,X}=-1.82\pm 0.04$.
The derived host galaxy extinction is $A_{R}=0.87$.
Attributing the early optical flare to the reverse-shock (RS) emission and the late optical-X-ray emission to the forward shock emission;
 the optical and X-ray lightcurves at $t<3\times 10^4$~s can be well fit adopting an Markov Chain Monte Carlo algorithm.
 Comparing the properties of GRB 210104A with other GRBs that have detection of bright RS emission,
 we show that its jet is mildly magnetized ($R_{\rm B}=28$), with high radiation efficiency ($77\%$),
 is sub-energetic ($E_{\rm k, iso}=4.5\times 10^{51}$ erg), and moderately relativistic ($\Gamma_0\sim 35$)
 in a density medium ($n_{0}\sim 417\;{\rm cm}^{-3}$). It follows the tight $L_{\gamma,\rm iso}-E_{\rm p,z}-\Gamma_{0}$ relation as with typical GRBs.
\end{abstract}


\section{Introduction}
\label{sec:introduction}
Gamma-ray bursts (GRBs) and their afterglows are the most luminous transients in the universe.
Their isotropic prompt gamma-ray energy is typically $10^{51}\sim 10^{54}$ erg, and their bright X-ray afterglows were detected with the X-Ray Telescope (XRT)
for almost $90\%$ of GRBs triggered with the Burst Alert Telescope (BAT) on board the Swift mission
(e.g., \citealp{2009ARA&A..47..567G}). Their optical afterglow can even can be seen by eyes.
The record holder so far is GRB 080319B, the so-called naked-eye GRB,
which has an extremely bright optical afterglow
with a peak visual magnitude of 5.3 at redshift of $z=0.937$
\citep{Racusin_Karpov_2008Natur.455..183R}.
Early bright optical flares were detected in some GRBs,
such as GRB 990123 ($\sim 9$ mag in the $R$ band at $z=1.6$;
 \citealp{Akerlof_Balsano_1999Natur.398..400A}),
GRB 080129 ($\sim 17.5$ mag in the $J$ band at $z=4.349\pm 0.002$; \citealp{Greiner_Kruhler_2009ApJ...693.1912G}),
GRB 090102 ($\sim 13.4$ in the $R$ band at $z=1.547$;
\citealp{Klotz_Gendre_2009GCN..8764....1K, Mangano_La_Parola_2009GCN..8767....1M}),
GRB 130427A ($7.03\pm 0.03$ mag in the $r^{\prime}$ band at $z=0.34$; \citealp{Vestrand_Vestra_nd2014Sci...343...38V}),
and GRB 140512A ($13$ mag in the $R$ band at $z=0.725$;
\citealp{Huang_Xin_2016ApJ...833..100H}).
These flare-like optical afterglows are thought
to be the external reverse-shock (RS) emission of the GRB jets
(e.g., \citealp{Meszaros_Rees_1999MNRAS.306L..39M, Sari_Piran_1999ApJ...520..641S,
Sari_Piran_1999ApJ...517L.109S,Kobayashi_Sari_2000ApJ...542..819K}).
They are powerful probes for investigating the physical properties
of the radiating region and its surrounding medium
(e.g., \citealp{Sari_Piran_1999ApJ...517L.109S, Kobayashi_Zhang_2003ApJ...597..455K,
Japelj_Kopavc_2014ApJ...785...84J,Gao_Wang_2015ApJ...810..160G, Yi_Wu_2020ApJ...895...94Y}).

Comparing to the onset peak of the forward shock (FS) emission,
the optical flares produced by the RS are rarely detected.
It was suggested that the RS emission could  be suppressed by strong magnetization of the RS region \citep{Zhang_Kobayashi_2005ApJ...628..315Z, Mimica_Giannios_2009A&A...494..879M, Mizuno_Zhang_2009ApJ...690L..47M}
or the RS emission peaks at a lower energy band
\citep{Mundell_Steele_2007Sci...315.1822M,
Melandri_Kobayashi_2010ApJ...723.1331M, Resmi_Zhang_2016ApJ...825...48R}.
Alternately, the RS emission may also be masked by the prompt emission from the internal shocks
\citep{Kopavc_Kobayashi_2013ApJ...772...73K}.
RS emission is a critical probe to study the magnetization of the jet
(e.g., \citealp{Chen_Liu_2021MNRAS.504.1759C}).
With detection of RS emission in some GRBs,
it is found that the magnetization of the RS region
is stronger than that in the FS region by a magnetization parameter of $R_{B}=\epsilon_{B}^{r}/\epsilon_{B}^{f}=2-10^4$,
where $\epsilon_{B}^{f}$ and $\epsilon_{B}^{r}$ are the magnetic equipartition parameters of the FS and RS regions \citep{Zhang_Kobayashi_2003ApJ...595..950Z, Japelj_Kopavc_2014ApJ...785...84J}.
The RS emission could dominate the early optical afterglow emission
if the typical value of $R_B\sim 100$ \citep{Gao_Wang_2015ApJ...810..160G}.

As a cosmic beacon, the GRB afterglows are probes for the ambient medium surrounding GRB jets (e.g., \citealp{2010AJ....139..694L, 2013MNRAS.432.1231C}). The medium density surrounding a short GRBs,
which are produced by the mergers of compact stars, is expected to be low
\citep{Panaitescu_Kumar_2001ApJ...561L.171P, Perna_Belczynski_2002ApJ...570..252P, 2011MNRAS.410...27X}.
Long GRBs originate from the collapse of massive stars
and they usually occur in the star formation region.
It is found that long GRB host galaxies generally have a low-metallicity interstellar medium (ISM) environment (e.g., \citealp{Fruchter_Levan_2006Natur.441..463F}).
The medium density profile of some GRBs can be parameterized as
$\rho \propto r^{-2}$, where $r$ is the distance to the progenitor,
implying the surrounding medium should be wind medium
(e.g., \citealp{Dai_Lu_1998MNRAS.298...87D, Meszaros_Rees_1998ApJ...499..301M,
Panaitescu_Meszaros_1998ApJ...503..314P,
Chevalier_Li_1999ApJ...520L..29C, Chevalier_Li_2000ApJ...536..195C}).
By analyzing the temporal feature of the optical afterglow onset bumps for a sample of GRBs, it is found that the medium density profile at the early stage is
$\rho \propto r^{-k}$ with a typical $k$ value as 1 (e.g., \citealp{Liang_Li_2013ApJ...774...13L}).
Some unusual temporal features,
such as a bump or the transition of a flux decaying slope,
may also hint at the unique properties of the ambient density,
e.g., a medium density jump or the transition of the medium profile,
or the preheated and ionized medium (e.g., \citealp{2002ApJ...565L..87D}; \citealp{2005ApJ...627..346B}; \citealp{2021ApJ...917...95Z}).

This paper reports our photometric and spectroscopic observations of GRB~210104A
with the GWAC-F60, 0.8~m TNT telescopes, and 2.16~m telescope at Xinglong observatory.
A bright optical flare was observed in its very early optical afterglows.
It is a valuable case for enlarging our GRB sample with detection of a very early optical flare for studying the GRB jet properties and ambient density.
Our observations and data reduction is reported in \S~{\ref{observations},
and the data analysis is present in \S~{\ref{data analysis}.
Discussion of our analysis results and implications for the GRB jet properties are reported in \S~{\ref{Dis}.
We summarize our results in \S~{\ref{summary}.
Throughout, the convention $Q=10^{n}Q_{n}$ in the cgs units and $F\propto t^{\alpha}\nu ^{\beta}$ are adopted.
We take the cosmology
parameters as $H_0=67.8~\rm km~s^{-1}~Mpc^{-1}$, and $\Omega_M=0.308$ \citep{Planck_Collaboration-2016-Ade-A&A...594A..13P}.

\section{Observations and Data Reduction}\label{observations}
At 11:26:59.87 UT on 2021 January 4 ,
the Fermi Gamma-Ray Burst Monitor (GBM) triggered and located GRB 210104A,
which was also detected by  Swift/BAT
\citep{ Malacaria_Fletcher_2021GCN.29246....1M, Troja_Bernardini_2021GCN.29233....1T}.
The Fermi/GBM final real-time localization is consistent with the Swift position \citep{Fermi_GBM_Team-2021GCN.29232....1F}.
Its X-ray afterglows were detected with Swift/XRT since $62.7$~s after the BAT trigger.
We download the Swift/XRT lightcurve data from the website
of Swift Burst Analyzer\footnote{
\url{https://www.swift.ac.uk/xrt_curves/01015873/}}.

We carried out photometric follow-up observations by both GWAC-F60A\footnote{
The GWAC-F60A telescope with a diameter of 60~cm is a follow-up facility
belonging to the GWAC system.}
and 0.8 m TNT telescopes\footnote{
TNT is a 0.8 m Tsinghua University - National Astronomical Observatory
of China Telescope at Xinglong Observatory runs by a custom-designed automation system
for GRB follow-up observations \citep{Zheng-2008-Deng-ChJAA...8..693Z}.
A PI $1300\times1340$ CCD and filters in the standard Johnson Bessel system
are equipped for TNT.}
located at Xinglong observatory,
National Astronomical Observatories of Chinese Academy of Sciences.
The follow-ups were started by GWAC-F60A and 0.8 m TNT
about 61 and 11246~s after the BAT trigger, respectively.
Note that optical afterglow observation with the Swift/UVOT began at 71~s after the BAT trigger \citep{Breeveld_Troja_2021GCN.29247....1B}. Our optical detection is earlier than UVOT. The standard Johnson-Cousins $B$, $R$ and $I$ bands
are used for GWAC-F60A, and the $B$ and $R$ bands for TNT.
The exposure times range from 10 to 2100~s
depending on the filters and the evolution of the brightness
\citep{Xin_Wang_2021GCN.29235....1X}.
Raw images taken by both telescopes were reduced
by following the standard routine in the \texttt{IRAF} package,
including bias, dark, flat-field corrections.
The standard aperture photometry was adopted for building the lightcurves,
in which the absolute calibration of the brightness was
performed by using the Sloan Digital Sky Survey (SDSS) catalog \citep{Adelman_McCarthy_Agueros_2008ApJS..175..297A},
with flux/mag conversion of the SDSS system into the Johnson-Cousins system\footnote{
\url{http://www.sdss.org/dr6/algorithms/sdssUBVRITransform.html\#Lupton2005}}.
Moreover, the $r$ band was observed by using the NEXT-0.6 m optical telescope
located at Nanshan, Xinjiang, China.
Observations started at 11:59:19 UT and ended at 22:34:07 UT on 2021 January 4  \citep{Zhu_Fu_2021GCN.29252....1Z}.
The observation log and the resulting absolutely calibrated magnitudes of
our photometric observations are listed in Table~\ref{Tab:opt-data}.

We carried out spectroscopic observations at 11:28:00 UT on 2021 Jan 4 ,
about $1$ hr after the burst trigger, with the 2.16 m telescope in Xinglong Observatory \citep{Fan_Wang_2016PASP..128k5005F}.
The optical spectrum of afterglow of the burst was taken by the
Beijing Faint Object Spectrograph and Camera (BFOSC) that is equipped
with a back-illuminated E2V55-30 AIMO CCD.
The grism G4 was used in the observation,
which provides a wavelength coverage of $3600-8400$ \AA\ in the observer frame.
A slit width of 1.8\arcsec\ oriented in the south-north direction was adopted.
The spectral resolution was resulted to be $10$\AA\ as measured from the sky lines.
The exposure time is 3600 seconds.
The wavelength calibration was carried out with the iron-argon comparison lamps,
and the flux calibrations with the observations of the Kitt Peak National Observatory standard stars \citep{Massey_Johnson_1998ApJ...505..793M}.
The observed two-dimensional spectrum was reduced by the standard procedures,
including bias subtraction and the flat-field correction using the {\tt IRAF} package.
The extracted one-dimensional spectrum was calibrated in wavelength
and in flux by the corresponding comparison lamp and standards.
The accuracy of the wavelength calibration is $\sim1$\AA.

Figure~{\ref{Spectrum}} shows the extracted spectrum from our spectroscopic observations
in the observer frame after smoothing by a box size of 3\AA.
Although the signal-to-noise ratio of the continuum is low,
one can potentially identify a few absorption features due to
\ion{Mg}{2} $\lambda$2800 and \ion{Fe}{2} multiplets.
This identification returns an estimation of redshift of $z=0.46$.

\section{Data Analysis}\label{data analysis}
We convert the magnitude of the optical data to the cgs units with the zero-point flux of each band from \cite{Bessell_Castelli_1998A&A...333..231B}.
Note that the optical data are corrected for the extinction of our Galaxy with
$R=0.091, I=0.063$, and $B=0.151$
\citep{Schlafly_Finkbeiner_2011ApJ...737..103S}.
We show the optical and X-ray afterglow lightcurves in the left panel of Figure~{\ref{Photometric}}. We empirically fit the lightcurves with a single power-law function or a smooth broken power-law (BPL) function
$F = F_0[{{({t}/{t_b})^{\alpha_1\omega}}+
({t}/{t_b})^{\alpha_2 \omega}} ]^{1/\omega}$, where
$\omega$ depicts the conjunction sharpness of the two segments. Note that the slopes derived from the smooth broken power-law model depend on $\omega$, especially for a very smooth, broad bump like the X-ray afterglow lightcurve. We fix $\omega$ as $1$ in our analysis.
We adopt the least-square algorithm for our fit and evaluate the goodness with the reduced $\chi^2$.

Our results are shown in the left panel of Figure~{\ref{Photometric}}.
The $\chi_r^2$ of our best fit to the $R$-band lightcurve is 1.74.
Our fit shows that the early $R$-band flare peaks at $t_{\rm p,O}=112\pm 7$~s,
and its rising and decaying slopes are $\alpha_{1, R}=2.02\pm 0.30$ and $\alpha_{2, R}=-2.07\pm 0.14$, respectively.
The slopes are similar to that usually seen in the early optical flares
(e.g., \citealp{Gao_Wang_2015ApJ...810..160G, Huang_Xin_2016ApJ...833..100H}).
The $R$-band lightcurve at $t>10^3$~s is dominated by another smooth BPL component,
and the decaying slope at the late epoch is $\alpha_{\rm 3,R}=-0.91\pm 0.03$.
The decaying phase of the early optical flare is also well observed in the $I$-band,
but the rising part and the peak are missed. By rescaling the flux level,
we found that the $R$-band lightcurve fit can apply to the $I$ band lightcurve,
i.e., $\alpha_{2, I}=-2.07$  and $\alpha_{3, I}=-0.91$, with $\chi^2=1.11$.
The afterglows in the $r$ and $B$ bands are detected only at $t>10^4$ seconds,
and their decay slopes are $\alpha_{3,r}=-0.76\pm 0.02$ ($\chi^2=2.95$) and $\alpha_{ 3,B}=-1.19\pm 0.06$ ($\chi^2=0.67$).

The early X-ray afterglow lightcurve is completely different from the optical afterglow. Following an initial rapid decaying phase, the lightcurve shows a very smooth bump.
The initial rapid decaying X-ray emission would be resulted from the prompt gamma-ray tail emission being due the curvature effect
(e.g. \citealp{2006ApJ...646..351L}). Ignoring this component, our fit gives $\alpha_{\rm 1,X}\sim 1$, $\alpha_{\rm 2,X}=-1.18\pm 0.01$, and $t_{\rm p, X}=313\pm 6$~s by fixing $\omega=1$. The $\chi_r^2$ of the fit is 1.45.

We make a joint spectral analysis of the optical and X-ray data in the time interval from $1.1\times10^{4}$ to $1.3\times10^{4}$~s.
Simultaneous observations in the X-ray, $R$, $I$, and $B$ bands are available in this selected time interval.
We extract the X-ray spectrum observed with Swift/XRT and adopt the \texttt{XSPEC} (version 12.12.1) package for making the spectral fit.
Extinction and \ion{H}{1} absorption of both our Galaxy and host galaxy are taken into account.
The Galactic \ion{H}{1} column density at the GRB direction is
$N_{\rm H}^{\rm Gal}= 5.09\times 10^{20}~{\rm cm}^{-2}$.
The extinction of our Galaxy in the GRB direction is
$E(B-V)= 0.04$ \citep{Schlegel_Finkbeiner_1998ApJ...500..525S}.
The extinction curve of the host galaxy is taken as the same as our Galaxy
\citep{Pei_1992ApJ...395..130P} by setting $R_V=3.08$.
We find that an absorbed power-law model is equated to fit the spectrum
and the derived photon index is $\Gamma_{\rm O,X}=1.82\pm0.04$.
The reduced $\chi^{2}$ is $39.34/40=0.98$.
The hydrogen absorption of the host galaxy is
$N_{\rm H}=(6.22\pm 2.31)\times10^{20}~{\rm cm}^{-2}$.
The color index of the host galaxy is
$E(B-V)=0.36\pm 0.06$,
inferring an extinction in the $R$ band of the host galaxy as
$A_R=0.87$~mag. This indicates that the burst environment is dusty.

\section{Theoretically Modeling Optical and X-Ray afterglow LightCurves}\label{modeling}
Our above analysis shows that the early optical flare seems to be attributed to the RS emission,
and the late optical and X-ray emission should could be the FS emission of the GRB fireball.
It is uncertain whether the early X-ray bump can be explained with the external shock model.
We fit the X-ray and optical lightcurves with the FS+RS afterglow model in this section.
The details of our FS model are present in
\cite{Sari_Piran_1998ApJ...497L..17S}, \cite{Huang_Gou_2000ApJ...543...90H}, \cite{Fan_Piran_2006MNRAS.369..197F}, and \cite{Ren_Lin_2020ApJ...901L..26R}.
The RS model is referred to
\cite{Yi_Wu_2013ApJ...776..120Y} and \cite{Gao_Wang_2015ApJ...810..160G}.
The model parameters are the initial bulk Lorentz factor ($\Gamma_{0}$),
the isotropic kinetic energy ($E_{\rm k, iso}$),
the medium number density ($n_0$),
the energy partitions of the electrons ($\epsilon_e^{f}$
and $\epsilon_e^{r}$)
and magnetic fields ($\epsilon_B^{f}$ and $\epsilon_B^{r}$)
as well as the electron energy distribution power-law indices ($p^{f}$ and $p^{r}$)
 of the FS and RS regions,
 where the superscripts ``$f$'' and ``$r$''
 indicate the parameters of the FS and RS regions, respectively.

We fit the $R$-band and X-ray afterglow lightcurves
with our model by adopting a Markov Chain Monte Carlo (MCMC) algorithm
available in the Python package {\tt emcee}
\citep{Foreman_Mackey_Hogg_2013PASP..125..306F}.
Note that the $R$-band lightcurve is corrected by the extinctions
of both the host galaxy and our Galaxy. The probability contours of the model parameters derived from out fit are shown in Figure~{\ref{corner}. The best fit yields the parameters and their $1\sigma$ uncertainty are:
${\rm log_{10}}\Gamma_{0}=1.55_{-0.02}^{+0.02}$,
${\rm log_{10}}E_{\rm k, iso}({\rm erg}) =51.65_{-0.02}^{+0.02}$,
${\rm log_{10}}n_0({\rm cm}^{-3})=2.62_{-0.15}^{+0.14}$,
${\rm log_{10}}\epsilon_e^{f}=-0.49_{-0.02}^{+0.02}$,
${\rm log_{10}}\epsilon_B^{f}=-3.90_{-0.13}^{+0.13}$,
$p^{f}=2.15_{-0.02}^{+0.02}$,
${\rm log_{10}}\epsilon_e^{r}=-0.84_{-0.05}^{+0.06}$,
${\rm log_{10}}\epsilon_B^{r}=-2.45_{-0.09}^{+0.10}$,
and $p^{r}=2.07_{-0.01}^{+0.01}$.

Our best fit is shown in Figure \ref{Model_fitting}.
It is found that the optical and X-ray afterglow lightcurves at $t<3\times 10^4$~s can be well fitted with our model.
However, the model predicted X-ray flux at $t>3\times 10^4$~s is higher than the observed one.
As shown in the left panel of Figure \ref{Photometric},
the decay slope of the late X-ray afterglow index \underline{}is $\alpha_{\rm 2, X}=1.18\pm 0.01$,
convincingly ruling out the possibility that the discrepancy results from the jet break effect.
Our joint spectral analysis for the optical-X-ray afterglows around $10^4$~s suggests that the X-ray and optical afterglows are in the same spectral regime, as shown in the right panel of Figure \ref{Photometric}.
It is possible that the late X-ray and optical afterglows are in different spectral regimes.
$\alpha_{\rm X}$ is steeper than that of the {\em R}-band afterglow ($\alpha_{3,R}=-0.91\pm 0.03$),
with a difference of $\Delta \alpha=0.27$. The flux decay slope of the synchrotron emission depends on the spectral regime (e.g., \citealp{Sari_Piran_1998ApJ...497L..17S, Zhang-2006-Fan-ApJ...642..354Z}).
In a constant-density medium, we have $\alpha=3(p^f-1)/4=0.86$ in the $\nu_{m}< \nu < \nu_{c}$ spectral regime and $\alpha=(3p^f-2)/4=1.1$ in the $\nu > \nu_{c}$ spectral regime by taking $p^f=2.15$.
Thus, the late optical afterglow should be in the $\nu_{m}< \nu < \nu_{c}$ spectral regime and the X-ray afterglows are in the $\nu_{\rm X} >\nu_c$ regime. Our best theoretical fit with the MCMC algorithm models the late X-ray and optical afterglows  in the same spectral regimes since the late X-ray data are spare and largely uncertain.

\section{Discussion}\label{Dis}
GRB 2101014A is of interest with its bright early optical flash.
As mentioned in \S~\ref{sec:introduction},
bright RS optical emission only detected in several GRBs (e.g., \citealp{Japelj_Kopavc_2014ApJ...785...84J, Gao_Wang_2015ApJ...810..160G}, and references therein).
The lack might be explained with the suppression by strong magnetization of the RS region \citep{Zhang_Kobayashi_2005ApJ...628..315Z, Mimica_Giannios_2009A&A...494..879M, Mizuno_Zhang_2009ApJ...690L..47M},
the RS emission peak at a lower frequency than the optical band \citep{Mundell_Steele_2007Sci...315.1822M,
Melandri_Kobayashi_2010ApJ...723.1331M, Resmi_Zhang_2016ApJ...825...48R},
and/or masked by the prompt optical and forward shock emission components
\citep{Kopavc_Kobayashi_2013ApJ...772...73K}.
Different from the bright optical flash observed in GRB 090727 \citep{Mundell_Steele_2007Sci...315.1822M},
the optical flash of GRB 210104A is not associated with prompt gamma-ray and/or X-ray flare.
It could not be originated from the internal shock.
We attributed it to the RS emission of the GRB jet in our analysis.
Detection of bright RS emission in the $R$ band indicates that the typical frequency ($\nu^r_{\rm m}$)
of the electron synchrotron emission in the RS region should be in the $R$ band.
We calculate the value of $\nu^{\rm r}_{\rm m}$ at the peak time of optical flash through
$\nu^{\rm r}_{\rm m}= 3.3 \times 10^{12}[(p^{\rm r}-2)/(p^{\rm r}-1)]^2 (1+z)^{1/2} {\epsilon^r_{\rm B,-2}}^{1/2}{\epsilon^r_{\rm e,-1}}^{2}{E_{\rm k, 52}}^{1/2}{t_{\rm p,day}}p^{-3/2}$ (e.g., \citealp{Sari_Piran_1998ApJ...497L..17S, Yost-2003-Harrison-ApJ...597..459Y}). We obtain $\nu^r_m(t_{\rm p})=2.86\times 10^{14}$~Hz, which is closed to the $R$ band $(4.28\times 10^{14}~\rm  Hz)$.

We compare the properties of GRB 210104A with a sample of GRBs available in \cite{Japelj_Kopavc_2014ApJ...785...84J}.
We derive the properties of GRB 210104A from the data observed with the GBM (8-1000 keV) on board the Fermi mission.
Its duration is $32$~s in the $50-300$~keV band \citep{Malacaria_Fletcher_2021GCN.29246....1M}.
We extract the GBM spectrum accumulated in the time interval
from $1.5$ to $33.5$~s since the GBM trigger with the tool
{\tt gtburst}\footnote{
The {\em Fermi}/GBM data are downloaded from the {\em Fermi} website
\url{https://heasarc.gsfc.nasa.gov/FTP/fermi/data/gbm/triggers/2021/bn210104477/current/}.}.
We use the data of two NaI detectors (na and nb)
and one BGO detector (b1) based on the angle between each detector and the source.
We fit the spectrum by using the Band function \citep{Band-1993-Matteson-ApJ...413..281B} with the {\tt XSPEC} package.
We obtain a low-energy photon index as $-1.09\pm 0.06$,
a high-energy photon index as $-2.63\pm 0.35$,
and a $\nu f_\nu$ spectrum peak energy $E_{\rm p}=199\pm 34$ {\rm keV}. The reduced $\chi^2$ of the fit is 474/343.
The derived $E_{\rm \gamma, \rm iso}$ in the $1-10^4$ keV band is $1.30^{+0.20}_{-0.23}\times10^{52}$~erg,
and its peak luminosity is $2.5\times 10^{51}$ erg~s$^{-1}$.
The prompt and afterglow emission properties, including $\Gamma_0$,
$E_{\rm \gamma, \rm iso}$, $L_{\rm \gamma,\rm iso}$, $E_{\rm k,\rm iso}$, and $E_{\rm p}$ of GRB 210104A together with
other GRBs from  \cite{Japelj_Kopavc_2014ApJ...785...84J}\footnote{GRBs 021004 and 060908 are not included in our comparison since they are detected with Swift/BAT only and no broadband observations are available for constraining its prompt gamma-ray spectrum.} are summarized in Table \ref{tab3}.
These properties are dramatically different from each other, and each parameter is in a broad range among these GRBs,
i.e., $E_{\rm \gamma, iso, 52}=1.1\sim 239$, $E_{\rm k, iso, 52}=0.39\sim 816$, $L_{\gamma, \rm iso,52}=0.25\sim 27$, and $E_{p}=47\sim 1028$ keV.
These results imply that these GRBs do not have any universal features of their prompt gamma-ray and afterglow emission.
They are not distinct from typical GRBs.

Measuring the magnetization of the jet with $R_{\rm B}\equiv  \epsilon_{\rm B}^{r}/\epsilon_{\rm B}^{f}$\footnote{$R_{\rm B}$ is defined as $R_{\rm B} \equiv (\epsilon_{\rm B}^{r}/\epsilon_{\rm B}^{f})^{1/2}$ in \cite{Zhang_Kobayashi_2003ApJ...595..950Z}.
Here we adopt the definition as $R_{\rm B}\equiv  \epsilon_{\rm B}^{r}/\epsilon_{\rm B}^{f}$.},
we calculate the $R_{\rm B}$ values of GRB 210104A and other GRBs in Table \ref{tab3}.
In addition, the radiation efficiency,
which is defined as
$\eta_{\gamma}\equiv E_{\rm \gamma,iso}/(E_{\rm \gamma,iso}+E_{\rm k,iso})$,
is also a key probe for the jet composition (e.g., \citealp{Zhang_Liang_2007ApJ...655..989Z, Wang_Zhang_2015ApJS..219....9W}).
We also calculate their gamma-ray radiation efficiencies.
Our results are also reported in Table \ref{tab3}. One can observe that $R_{\rm B}$ and $\eta_\gamma$ values are dramatically different among these GRBs,
i.e. $R_{\rm B}=4\sim 16540$ and $\eta_\gamma=0.02\sim 0.77$. The radiation efficiency of GRB 210104A is the largest one among these GRBs, i.e., $\eta_\gamma=77\%$.
The high radiation efficiency is consistent with the expectation of a magnetized jet
(e.g., \citealp{Zhang_Liang_2007ApJ...655..989Z, Wang_Zhang_2015ApJS..219....9W}).
\cite{Gao_Wang_2015ApJ...810..160G} reported a typical $R_{\rm B}$ value of $\sim 100$ for a sample of GRBs with detection of RS optical emission.
GRB 210104A has a relative low $R_{\rm B}$ value among these GRBs.  Its $R_{\rm B}$ value is 28,
indicating that its jet is only moderately magnetized.

The derived $\Gamma_0$ of GRB 210104A from our fit is 35,
indicating that the jet is middle relativistic.
We test whether such a $\Gamma_0$ value violates the lower limit derived from the pair production opacity constraint.
The spectrum of GRB 2101014A can be fitted with the Band function.
Its high-energy spectrum is $f(E)=f(E_{c})(E/E_{c})^{\beta}$ for $E>E_{c}$, where $E_{c}=(\alpha-\beta)E_{\rm peak}/(2+\alpha)$ and $f(E_c)=A \exp(\beta-\alpha)(E_c/100~\rm {keV})^{\alpha}$.
The minimum bulk Lorentz factor for the detection of maximum photon with energy $E_0$  sets a lower limit of $\Gamma_0$ as  (e.g., \citealp{Lithwick-2001-Sari-ApJ...555..540L}; see also \citealp{Abdo_2009-Ackermann-Sci...323.1688A})

\begin{equation}
\Gamma_{\min }=\left[\sigma_{T}\left(\frac{d_{L}(z)}{c \Delta t}\right)^{2} E_{c} f\left(E_{c}\right) F(\beta)\right]^{\frac{1}{2(1-\beta)}}(1+z)^{\frac{\beta+1}{1-\beta}}\left(\frac{E_{0} E_{c}}{m_{e}^{2} c^{4}}\right)^{\frac{\beta+1}{2(\beta-1)}},
\end{equation}

where $\sigma_{\rm T}$ is the Thompson scattering area, $c$ is speed of light, $d_{\rm L}$ is the luminosity distance, $m_{\rm e}$ is the electron mass,
$F(\beta)$ is a dimensionless function $F(\beta)=4/(1-\beta)\int_{0}^{1}dy(1-y^{2})^{-\beta-2}g(y)y$,
where $g(y)=3/16(1-y^{2})\{(3-y^{4})\ln[(1+y)/(1-y)]-2y(2-y^{2})\}$,
and $\Delta t$ is the minimum variation timescale of prompt emission, which is taken as a typical value of $0.1$~s \citep{MacLachlan-2013-Shenoy-MNRAS.432..857M}.
The Konus-Wind mission detected the highest photon energy up to $2~\rm MeV$ from GRB 210104A \citep{Frederiks-2021-Golenetskii-GCN.29258....1F}.
Thus, we have $\Gamma_{\rm min}=33$.
The $\Gamma_0$ value derived from our fit is close to $\Gamma_{\rm min}$,
but still does not violate the pair production opacity constraint.

\cite{Liang_Lin_2015ApJ...813..116L} found a tight
$L_{\gamma,\rm iso}-E_{p,z}-\Gamma_0$ correlation,
i.e., $L^{\rm r}_{\rm \gamma, iso}=10^{45.62\pm 0.35} ~{\rm erg~s^{-1}}~
{(E_{p,z}/{\rm keV})}^{1.34\pm 0.14}\Gamma_{0}^{1.32\pm 0.19}$, where $E_{\rm p,z}=E_{\rm p}(1+z)$. This relation combines the jet luminosity (or energy), the initial Lorentz factor, and the radiation spectrum. It significantly reduces the intrinsic scatters of the $L_{\gamma,\rm iso}-E_{p,z}$
\citep{Liang_Dai_2004ApJ...606L..29L, Amati_2006MNRAS.372..233A}
and $L_{\gamma,\rm iso}-\Gamma_0$
\citep{Liang_Yi_2010ApJ...725.2209L, Lu_Zou_2012ApJ...751...49L} relations, resulting in a much tighter relations than the $L_{\gamma,\rm iso}-E_{\rm p,z}$ and $L_{\gamma,\rm iso}-\Gamma_0$ relation. This relation, which combines the observed luminosity with the properties of the GRB central engine,
the medium density, and the radiation physics, should be an intrinsic feature of GRB jets
(see also \citealp{Huang_Liang_2020ApJ...903L..26H}).
We examine whether GRB 210104A and other GRBs in Table \ref{tab3} satisfy the $L_{\gamma,\rm iso}-E_{p,z}-\Gamma_0$ correlation.
We calculate the $L^{\rm r}_{p,\rm iso, 52}$  values and show $L^{\rm r}_{p,\rm iso, 52}$ as a function of $L_{p,\rm iso, 52}$ in Figure \ref{Correlation}.
It is found that these GRBs follow well this relation.
The RS GRBs taken from  \cite{Japelj_Kopavc_2014ApJ...785...84J} tend to locate at the high luminosity end of this relation,
indicating that their prompt gamma-ray luminosities are averagely brighter than typical LGRBs.
This would be due to the sample selection effect. GRB 210104A is a relatively low-luminosity, subenergetic GRB among the RS GRBs.

\section{Summary}
\label{summary}
We have reported our spectroscopic and photometric observations of GRB 210104A
and present our analysis on its properties of the jet and ambient medium.
Our results are summarized below.
\begin{itemize}
\item  The afterglow spectrum obtained from our spectroscopic observations
at about 1 hr after the burst trigger reveals several absorption line features
due to absorptions by \ion{Mg}{2} $\lambda$ 2800 and \ion{Fe}{2},
and the redshift is estimated as $z = 0.46$.

\item A bright early optical flare is observed in the $R$ band in the time interval $67\sim 152$~s post the GRB trigger.
The flash has a peak brightness of $R=13$ mag at $112\pm 7$~s.
The slope of the $R$-band afterglow lightcurve is $\alpha_{\rm 3, R}={-0.91\pm 0.03}$ at $t>650$~s.
The X-ray afterglow lightcurve features as a smooth onset bump at the early epoch and decays with a slope of $\alpha_{\rm 2,X}=-1.18\pm 0.01$.
Our joint optical-X-ray spectral analysis in the time slice of
$t\in; [1.1, 1.3]\times10^{4}$~s shows that a power-law
with a photon index of $\Gamma_{\rm O,X}=1.82\pm0.04$
is equated to fit the spectrum. The derived \ion{H}{1} absorption and dust extinction of the host galaxy
are $N_{\rm H}=6.22_{-2.31}^{+2.31}\times10^{20}~{\rm cm}^{-2}$
and $A_{\rm R}=0.87$~mag by adopting an extinction curve as our Galaxy.

\item  The $R$-band and X-ray afterglow lightcurves of GRB 210104A at $t<3\times 10^4$~s
can be fitted with the standard RS and FS models by adopting an MCMC algorithm.
The early bright optical flare is attributed to the RS emission of the jet.
The best fit yields the parameters and their $1\sigma$ uncertainties are:
${\rm log_{10}}\Gamma_{0}=1.55_{-0.02}^{+0.02}$,
${\rm log_{10}}E_{\rm k, iso}({\rm erg}) =51.65_{-0.02}^{+0.02}$,
${\rm log_{10}}n_0({\rm cm}^{-3})=2.62_{-0.15}^{+0.14}$,
${\rm log_{10}}\epsilon_e^{f}=-0.49_{-0.02}^{+0.02}$,
${\rm log_{10}}\epsilon_B^{f}=-3.90_{-0.13}^{+0.13}$,
$p^{f}=2.15_{-0.02}^{+0.02}$,
${\rm log_{10}}\epsilon_e^{r}=-0.84_{-0.05}^{+0.06}$,
${\rm log_{10}}\epsilon_B^{r}=-2.45_{-0.09}^{+0.10}$,
and $p^{r}=2.07_{-0.01}^{+0.01}$.
The derived $\nu^{\rm r}_m$ is very closed to the $R$ band.
Our theoretical analysis shows that the RS region is mildly magnetized ($R_{\rm B}=28$),
suggesting that the jet is only moderately magnetized. The gamma-ray radiating efficiency is 77\%.
The jet is moderately relativistic ($\Gamma_0=35$).
Such a low $\Gamma_0$ value does not violate the the lower limit derived from the pair production opacity constraint ( $\Gamma_{\rm min}=33$).

\item We compare the properties of GRB 210104A with a sample of GRBs whose RS emission is detected in their early optical afterglows from  \cite{Japelj_Kopavc_2014ApJ...785...84J}.
We do not find any universal features of their prompt gamma-ray and forward shock emission.
They are not distinct from typical GRBs. They follow the $L_{\gamma, \rm iso}-E_{\rm p,z}-\Gamma_0$ relation derived from typical GRBs.
\end{itemize}

\acknowledgments
We would like to thank the referee for a constructive report and thank Jia Ren for helpful discussions.
We acknowledge the use of the public data from the
Swift data archive and the UK Swift Science Data Center.
This work is supported by the National Natural Science Foundation of China
(grant Nos. 12133003, 11851304,
and U1731239, 11973055, U1831207, 11863007, U1931133, U1938201).
We acknowledge the support of the staff of the Xinglong 2.16 m telescope.
This work was partially supported by the Open Project Program of the Key Laboratory of Optical Astronomy, National Astronomical Observatories, Chinese Academy of Sciences.

\facilities{\em GWAC-F60A and NAOC 2.16m optical telescopes at Xinglong Observatory, Heibei, China; NEXT-0.6m optical telescope located at Nanshan, Xinjiang, China}

\software{\texttt{IRAF} \citep{Tody_1986SPIE..627..733T,Tody_1992_Davis_ASPC...25..484T},
\texttt{emcee}\citep{Foreman_Mackey_Hogg_2013PASP..125..306F}}

\clearpage


\begin{thebibliography}{}
\expandafter\ifx\csname natexlab\endcsname\relax\def\natexlab#1{#1}\fi
\providecommand{\url}[1]{\href{#1}{#1}}
\providecommand{\dodoi}[1]{doi:~\href{http://doi.org/#1}{\nolinkurl{#1}}}
\providecommand{\doeprint}[1]{\href{http://ascl.net/#1}{\nolinkurl{http://ascl.net/#1}}}
\providecommand{\doarXiv}[1]{\href{https://arxiv.org/abs/#1}{\nolinkurl{https://arxiv.org/abs/#1}}}

\bibitem[{{Abdo} {et~al.}(2009){Abdo}, {Ackermann}, {Arimoto}, {Asano},
  {Atwood}, {Axelsson}, {Baldini}, {Ballet}, {Band}, {Barbiellini}, {Baring},
  {Bastieri}, {Battelino}, {Baughman}, {Bechtol}, {Bellardi}, {Bellazzini},
  {Berenji}, {Bhat}, {Bissaldi}, {Blandford}, {Bloom}, {Bogaert}, {Bogart},
  {Bonamente}, {Bonnell}, {Borgland}, {Bouvier}, {Bregeon}, {Brez}, {Briggs},
  {Brigida}, {Bruel}, {Burnett}, {Burrows}, {Busetto}, {Caliandro}, {Cameron},
  {Caraveo}, {Casandjian}, {Ceccanti}, {Cecchi}, {Celotti}, {Charles},
  {Chekhtman}, {Cheung}, {Chiang}, {Ciprini}, {Claus}, {Cohen-Tanugi},
  {Cominsky}, {Connaughton}, {Conrad}, {Costamante}, {Cutini}, {DeKlotz},
  {Dermer}, {de Angelis}, {de Palma}, {Digel}, {Dingus}, {do Couto e Silva},
  {Drell}, {Dubois}, {Dumora}, {Edmonds}, {Evans}, {Fabiani}, {Farnier},
  {Favuzzi}, {Finke}, {Fishman}, {Focke}, {Frailis}, {Fukazawa}, {Funk},
  {Fusco}, {Gargano}, {Gasparrini}, {Gehrels}, {Germani}, {Giebels},
  {Giglietto}, {Giommi}, {Giordano}, {Glanzman}, {Godfrey}, {Goldstein},
  {Granot}, {Greiner}, {Grenier}, {Grondin}, {Grove}, {Guillemot}, {Guiriec},
  {Haller}, {Hanabata}, {Harding}, {Hayashida}, {Hays}, {Morata}, {Hoover},
  {Hughes}, {J{\'o}hannesson}, {Johnson}, {Johnson}, {Johnson}, {Johnson},
  {Kamae}, {Katagiri}, {Kataoka}, {Kavelaars}, {Kawai}, {Kelly}, {Kennea},
  {Kerr}, {Kippen}, {Kn{\"o}dlseder}, {Kocevski}, {Kocian}, {Komin},
  {Kouveliotou}, {Kuehn}, {Kuss}, {Lande}, {Landriu}, {Larsson}, {Latronico},
  {Lavalley}, {Lee}, {Lee}, {Lemoine-Goumard}, {Lichti}, {Longo}, {Loparco},
  {Lott}, {Lovellette}, {Lubrano}, {Madejski}, {Makeev}, {Marangelli},
  {Mazziotta}, {McBreen}, {McEnery}, {McGlynn}, {Meegan}, {M{\'e}sz{\'a}ros},
  {Meurer}, {Michelson}, {Minuti}, {Mirizzi}, {Mitthumsiri}, {Mizuno},
  {Moiseev}, {Monte}, {Monzani}, {Moretti}, {Morselli}, {Moskalenko}, {Murgia},
  {Nakamori}, {Nelson}, {Nolan}, {Norris}, {Nuss}, {Ohno}, {Ohsugi}, {Okumura},
  {Omodei}, {Orlando}, {Ormes}, {Ozaki}, {Paciesas}, {Paneque}, {Panetta},
  {Parent}, {Pelassa}, {Pepe}, {Perri}, {Pesce-Rollins}, {Petrosian},
  {Pinchera}, {Piron}, {Porter}, {Preece}, {Rain{\`o}}, {Ramirez-Ruiz},
  {Rando}, {Rapposelli}, {Razzano}, {Razzaque}, {Rea}, {Reimer}, {Reimer},
  {Reposeur}, {Reyes}, {Ritz}, {Rochester}, {Rodriguez}, {Roth}, {Ryde},
  {Sadrozinski}, {Sanchez}, {Sander}, {Parkinson}, {Scargle}, {Schalk},
  {Segal}, {Sgr{\`o}}, {Shimokawabe}, {Siskind}, {Smith}, {Smith}, {Spandre},
  {Spinelli}, {Stamatikos}, {Starck}, {Stecker}, {Steinle}, {Stephens},
  {Strickman}, {Suson}, {Tagliaferri}, {Tajima}, {Takahashi}, {Takahashi},
  {Tanaka}, {Tenze}, {Thayer}, {Thayer}, {Thompson}, {Tibaldo}, {Torres},
  {Tosti}, {Tramacere}, {Turri}, {Tuvi}, {Usher}, {van der Horst}, {Vigiani},
  {Vilchez}, {Vitale}, {von Kienlin}, {Waite}, {Williams}, {Wilson-Hodge},
  {Winer}, {Wood}, {Wu}, {Yamazaki}, {Ylinen}, {Ziegler}, {Fermi LAT
  Collaboration}, \& {Fermi GBM
  Collaboration}}]{Abdo_2009-Ackermann-Sci...323.1688A}
{Abdo}, A.~A., {Ackermann}, M., {Arimoto}, M., {et~al.} 2009, Science, 323,
  1688, \dodoi{10.1126/science.1169101}

\bibitem[{{Adelman-McCarthy} {et~al.}(2008){Adelmpan-McCarthy}, {Ag{\"u}eros},
  {Allam}, {Allende Prieto}, {Anderson}, {Anderson}, {Annis}, {Bahcall},
  {Bailer-Jones}, {Baldry}, {Barentine}, {Bassett}, {Becker}, {Beers}, {Bell},
  {Berlind}, {Bernardi}, {Blanton}, {Bochanski}, {Boroski}, {Brinchmann},
  {Brinkmann}, {Brunner}, {Budav{\'a}ri}, {Carliles}, {Carr}, {Castander},
  {Cinabro}, {Cool}, {Covey}, {Csabai}, {Cunha}, {Davenport}, {Dilday}, {Doi},
  {Eisenstein}, {Evans}, {Fan}, {Finkbeiner}, {Friedman}, {Frieman},
  {Fukugita}, {G{\"a}nsicke}, {Gates}, {Gillespie}, {Glazebrook}, {Gray},
  {Grebel}, {Gunn}, {Gurbani}, {Hall}, {Harding}, {Harvanek}, {Hawley},
  {Hayes}, {Heckman}, {Hendry}, {Hindsley}, {Hirata}, {Hogan}, {Hogg}, {Hyde},
  {Ichikawa}, {Ivezi{\'c}}, {Jester}, {Johnson}, {Jorgensen}, {Juri{\'c}},
  {Kent}, {Kessler}, {Kleinman}, {Knapp}, {Kron}, {Krzesinski}, {Kuropatkin},
  {Lamb}, {Lampeitl}, {Lebedeva}, {Lee}, {French Leger}, {L{\'e}pine}, {Lima},
  {Lin}, {Long}, {Loomis}, {Loveday}, {Lupton}, {Malanushenko}, {Malanushenko},
  {Mandelbaum}, {Margon}, {Marriner}, {Mart{\'\i}nez-Delgado}, {Matsubara},
  {McGehee}, {McKay}, {Meiksin}, {Morrison}, {Munn}, {Nakajima}, {Neilsen},
  {Newberg}, {Nichol}, {Nicinski}, {Nieto-Santisteban}, {Nitta}, {Okamura},
  {Owen}, {Oyaizu}, {Padmanabhan}, {Pan}, {Park}, {Peoples}, {Pier}, {Pope},
  {Purger}, {Raddick}, {Re Fiorentin}, {Richards}, {Richmond}, {Riess}, {Rix},
  {Rockosi}, {Sako}, {Schlegel}, {Schneider}, {Schreiber}, {Schwope}, {Seljak},
  {Sesar}, {Sheldon}, {Shimasaku}, {Sivarani}, {Allyn Smith}, {Snedden},
  {Steinmetz}, {Strauss}, {SubbaRao}, {Suto}, {Szalay}, {Szapudi}, {Szkody},
  {Tegmark}, {Thakar}, {Tremonti}, {Tucker}, {Uomoto}, {Vanden Berk},
  {Vandenberg}, {Vidrih}, {Vogeley}, {Voges}, {Vogt}, {Wadadekar}, {Weinberg},
  {West}, {White}, {Wilhite}, {Yanny}, {Yocum}, {York}, {Zehavi}, \&
  {Zucker}}]{Adelman_McCarthy_Agueros_2008ApJS..175..297A}
{Adelman-McCarthy}, J.~K., {Ag{\"u}eros}, M.~A., {Allam}, S.~S., {et~al.} 2008,
  \apjs, 175, 297, \dodoi{10.1086/524984}

\bibitem[{{Akerlof} {et~al.}(1999){Akerlof}, {Balsano}, {Barthelmy}, {Bloch},
  {Butterworth}, {Casperson}, {Cline}, {Fletcher}, {Frontera}, {Gisler},
  {Heise}, {Hills}, {Kehoe}, {Lee}, {Marshall}, {McKay}, {Miller}, {Piro},
  {Priedhorsky}, {Szymanski}, \& {Wren}}]{Akerlof_Balsano_1999Natur.398..400A}
{Akerlof}, C., {Balsano}, R., {Barthelmy}, S., {et~al.} 1999, \nat, 398, 400,
  \dodoi{10.1038/18837}

\bibitem[{{Amati}(2006)}]{Amati_2006MNRAS.372..233A}
{Amati}, L. 2006, \mnras, 372, 233, \dodoi{10.1111/j.1365-2966.2006.10840.x}

\bibitem[{{Band} {et~al.}(1993){Band}, {Matteson}, {Ford}, {Schaefer},
  {Palmer}, {Teegarden}, {Cline}, {Briggs}, {Paciesas}, {Pendleton}, {Fishman},
  {Kouveliotou}, {Meegan}, {Wilson}, \&
  {Lestrade}}]{Band-1993-Matteson-ApJ...413..281B}
{Band}, D., {Matteson}, J., {Ford}, L., {et~al.} 1993, \apj, 413, 281,
  \dodoi{10.1086/172995}

\bibitem[{{Bellm} {et~al.}(2006){Bellm}, {Bandstra}, {Boggs}, {Wigger},
  {Hajdas}, {Smith}, \& {Hurley}}]{Bellm-2006-Bandstra-GCN..5867....1B}
{Bellm}, E., {Bandstra}, M., {Boggs}, S., {et~al.} 2006, GRB Coordinates
  Network, 5867, 1

\bibitem[Beloborodov(2005)]{2005ApJ...627..346B}
 Beloborodov, A.~M.\ 2005, \apj, 627, 346. \dodoi{10.1086/430166}p

\bibitem[{{Bessell} {et~al.}(1998){Bessell}, {Castelli}, \&
  {Plez}}]{Bessell_Castelli_1998A&A...333..231B}
{Bessell}, M.~S., {Castelli}, F., \& {Plez}, B. 1998, \aap, 333, 231

\bibitem[{{Breeveld} {et~al.}(2021){Breeveld}, {Troja}, \& {Swift/UVOT
  Team}}]{Breeveld_Troja_2021GCN.29247....1B}
{Breeveld}, A.~A., {Troja}, E., \& {Swift/UVOT Team}. 2021, GRB Coordinates
  Network, 29247, 1

\bibitem[{{Briggs} {et~al.}(1999){Briggs}, {Band}, {Kippen}, {Preece},
  {Kouveliotou}, {van Paradijs}, {Share}, {Murphy}, {Matz}, {Connors},
  {Winkler}, {McConnell}, {Ryan}, {Williams}, {Young}, {Dingus}, {Catelli}, \&
  {Wijers}}]{Briggs-1999-Band-ApJ...524...82B}
{Briggs}, M.~S., {Band}, D.~L., {Kippen}, R.~M., {et~al.} 1999, \apj, 524, 82,
  \dodoi{10.1086/307808}

\bibitem[{{Chen} \& {Liu}(2021)}]{Chen_Liu_2021MNRAS.504.1759C}
{Chen}, Q., \& {Liu}, X.-W. 2021, \mnras, 504, 1759,
  \dodoi{10.1093/mnras/stab946}

\bibitem[{{Chevalier} \& {Li}(1999)}]{Chevalier_Li_1999ApJ...520L..29C}
{Chevalier}, R.~A., \& {Li}, Z.-Y. 1999, \apjl, 520, L29,
  \dodoi{10.1086/312147}

\bibitem[{{Chevalier} \& {Li}(2000)}]{Chevalier_Li_2000ApJ...536..195C}
---. 2000, \apj, 536, 195, \dodoi{10.1086/308914}

\bibitem[Covino et al.(2013)]{2013MNRAS.432.1231C}
Covino, S., Melandri, A., Salvaterra, R., et al.\ 2013,
\mnras, 432, 1231. \dodoi{10.1093/mnras/stt540}

\bibitem[Dai \& Lu(2002)]{2002ApJ...565L..87D}
Dai, Z.~G. \& Lu, T.\ 2002, \apjl, 565, L87.
\dodoi{10.1086/339418}p


\bibitem[{{Dai} \& {Lu}(1998)}]{Dai_Lu_1998MNRAS.298...87D}
{Dai}, Z.~G., \& {Lu}, T. 1998, \mnras, 298, 87,
  \dodoi{10.1046/j.1365-8711.1998.01681.x}

\bibitem[{{Fan} \& {Piran}(2006)}]{Fan_Piran_2006MNRAS.369..197F}
{Fan}, Y., \& {Piran}, T. 2006, \mnras, 369, 197,
  \dodoi{10.1111/j.1365-2966.2006.10280.x}


\bibitem[{{Fan} {et~al.}(2016){Fan}, {Wang}, {Jiang}, {Wu}, {Li}, {Huang},
  {Xu}, {Hu}, {Zhu}, {Wang}, {Komossa}, \&
  {Zhang}}]{Fan_Wang_2016PASP..128k5005F}
{Fan}, Z., {Wang}, H., {Jiang}, X., {et~al.} 2016, \pasp, 128, 115005,
  \dodoi{10.1088/1538-3873/128/969/115005}

\bibitem[{{Fermi GBM Team}(2021)}]{Fermi_GBM_Team-2021GCN.29232....1F}
{Fermi GBM Team}. 2021, GRB Coordinates Network, 29232, 1


\bibitem[{{Foreman-Mackey} {et~al.}(2013){Foreman-Mackey}, {Hogg}, {Lang}, \&
  {Goodman}}]{Foreman_Mackey_Hogg_2013PASP..125..306F}
{Foreman-Mackey}, D., {Hogg}, D.~W., {Lang}, D., \& {Goodman}, J. 2013, \pasp,
  125, 306, \dodoi{10.1086/670067}

 \bibitem[{{Frederiks} {et~al.}(2021){Frederiks}, {Golenetskii}, {Lysenko},
  {Ridnaia}, {Svinkin}, {Tsvetkova}, {Ulanov}, {Cline}, \& {Konus-Wind
  Team}}]{Frederiks-2021-Golenetskii-GCN.29258....1F}
{Frederiks}, D., {Golenetskii}, S., {Lysenko}, A., {et~al.} 2021, GRB
  Coordinates Network, 29258, 1

\bibitem[{{Fruchter} {et~al.}(2006){Fruchter}, {Levan}, {Strolger},
  {Vreeswijk}, {Thorsett}, {Bersier}, {Burud}, {Castro Cer{\'o}n},
  {Castro-Tirado}, {Conselice}, {Dahlen}, {Ferguson}, {Fynbo}, {Garnavich},
  {Gibbons}, {Gorosabel}, {Gull}, {Hjorth}, {Holland}, {Kouveliotou}, {Levay},
  {Livio}, {Metzger}, {Nugent}, {Petro}, {Pian}, {Rhoads}, {Riess}, {Sahu},
  {Smette}, {Tanvir}, {Wijers}, \&
  {Woosley}}]{Fruchter_Levan_2006Natur.441..463F}
{Fruchter}, A.~S., {Levan}, A.~J., {Strolger}, L., {et~al.} 2006, \nat, 441,
  463, \dodoi{10.1038/nature04787}

\bibitem[{{Gao} {et~al.}(2015){Gao}, {Wang}, {M{\'e}sz{\'a}ros}, \&
  {Zhang}}]{Gao_Wang_2015ApJ...810..160G}
{Gao}, H., {Wang}, X.-G., {M{\'e}sz{\'a}ros}, P., \& {Zhang}, B. 2015, \apj,
  810, 160, \dodoi{10.1088/0004-637X/810/2/160}

\bibitem[Gehrels et al.(2009)]{2009ARA&A..47..567G}
Gehrels, N., Ramirez-Ruiz, E., \& Fox, D.~B.\ 2009,
\araa, 47, 567. \dodoi{10.1146/annurev.astro.46.060407.145147}

\bibitem[{{Golenetskii} {et~al.}(2009){Golenetskii}, {Aptekar}, {Mazets},
  {Pal'Shin}, {Frederiks}, {Oleynik}, {Ulanov}, {Svinkin}, \&
  {Cline}}]{Golenetskii_2009_Aptekar-GCN..8776....1G}
{Golenetskii}, S., {Aptekar}, R., {Mazets}, E., {et~al.} 2009, GRB Coordinates
  Network, 8776, 1

\bibitem[{{Greiner} {et~al.}(2009){Greiner}, {Kr{\"u}hler}, {McBreen},
  {Ajello}, {Giannios}, {Schwarz}, {Savaglio}, {Yolda{\c{s}}}, {Clemens},
  {Stefanescu}, {Sala}, {Bertoldi}, {Szokoly}, \&
  {Klose}}]{Greiner_Kruhler_2009ApJ...693.1912G}
{Greiner}, J., {Kr{\"u}hler}, T., {McBreen}, S., {et~al.} 2009, \apj, 693,
  1912, \dodoi{10.1088/0004-637X/693/2/1912}

\bibitem[{{Guidorzi} {et~al.}(2005){Guidorzi}, {Frontera}, {Montanari},
  {Rossi}, {Amati}, {Gomboc}, {Hurley}, \&
  {Mundell}}]{Guidorzi-2005-Frontera-MNRAS.363..315G}
{Guidorzi}, C., {Frontera}, F., {Montanari}, E., {et~al.} 2005, \mnras, 363,
  315, \dodoi{10.1111/j.1365-2966.2005.09450.x}

\bibitem[{{Huang} {et~al.}(2020){Huang}, {Liang}, {Liu}, {Cheng}, \&
  {Wang}}]{Huang_Liang_2020ApJ...903L..26H}
{Huang}, X.-L., {Liang}, E.-W., {Liu}, R.-Y., {Cheng}, J.-G., \& {Wang}, X.-Y.
  2020, \apjl, 903, L26, \dodoi{10.3847/2041-8213/abc330}

\bibitem[{{Huang} {et~al.}(2016){Huang}, {Xin}, {Yi}, {Zhong}, {Qiu}, {Deng},
  {Wei}, \& {Liang}}]{Huang_Xin_2016ApJ...833..100H}
{Huang}, X.-L., {Xin}, L.-P., {Yi}, S.-X., {et~al.} 2016, \apj, 833, 100,
  \dodoi{10.3847/1538-4357/833/1/100}

\bibitem[{{Huang} {et~al.}(2000){Huang}, {Gou}, {Dai}, \&
  {Lu}}]{Huang_Gou_2000ApJ...543...90H}
{Huang}, Y.~F., {Gou}, L.~J., {Dai}, Z.~G., \& {Lu}, T. 2000, \apj, 543, 90,
  \dodoi{10.1086/317076}

\bibitem[{{Japelj} {et~al.}(2014){Japelj}, {Kopa{\v{c}}}, {Kobayashi},
  {Harrison}, {Guidorzi}, {Virgili}, {Mundell}, {Melandri}, \&
  {Gomboc}}]{Japelj_Kopavc_2014ApJ...785...84J}
{Japelj}, J., {Kopa{\v{c}}}, D., {Kobayashi}, S., {et~al.} 2014, \apj, 785, 84,
  \dodoi{10.1088/0004-637X/785/2/84}

\bibitem[{{Jin} {et~al.}(2013){Jin}, {Covino}, {Della Valle}, {Ferrero},
  {Fugazza}, {Malesani}, {Melandri}, {Pian}, {Salvaterra}, {Bersier},
  {Campana}, {Cano}, {Castro-Tirado}, {D'Avanzo}, {Fynbo}, {Gomboc},
  {Gorosabel}, {Guidorzi}, {Haislip}, {Hjorth}, {Kobayashi}, {LaCluyze},
  {Marconi}, {Mazzali}, {Mundell}, {Piranomonte}, {Reichart},
  {S{\'a}nchez-Ram{\'\i}rez}, {Smith}, {Steele}, {Tagliaferri}, {Tanvir},
  {Valenti}, {Vergani}, {Vestrand}, {Walker}, \&
  {Wo{\'z}niak}}]{Jin-2013-Covino-ApJ...774..114J}
{Jin}, Z.-P., {Covino}, S., {Della Valle}, M., {et~al.} 2013, \apj, 774, 114,
  \dodoi{10.1088/0004-637X/774/2/114}

\bibitem[{{Klotz} {et~al.}(2009){Klotz}, {Gendre}, {Boer}, \&
  {Atteia}}]{Klotz_Gendre_2009GCN..8764....1K}
{Klotz}, A., {Gendre}, B., {Boer}, M., \& {Atteia}, J.~L. 2009, GRB Coordinates
  Network, 8764, 1

\bibitem[{{Kobayashi} \& {Sari}(2000)}]{Kobayashi_Sari_2000ApJ...542..819K}
{Kobayashi}, S., \& {Sari}, R. 2000, \apj, 542, 819, \dodoi{10.1086/317021}

\bibitem[{{Kobayashi} \& {Zhang}(2003)}]{Kobayashi_Zhang_2003ApJ...597..455K}
{Kobayashi}, S., \& {Zhang}, B. 2003, \apj, 597, 455, \dodoi{10.1086/378283}

\bibitem[{{Kopa{\v{c}}} {et~al.}(2013){Kopa{\v{c}}}, {Kobayashi}, {Gomboc},
  {Japelj}, {Mundell}, {Guidorzi}, {Melandri}, {Bersier}, {Cano}, {Smith},
  {Steele}, \& {Virgili}}]{Kopavc_Kobayashi_2013ApJ...772...73K}
{Kopa{\v{c}}}, D., {Kobayashi}, S., {Gomboc}, A., {et~al.} 2013, \apj, 772, 73,
  \dodoi{10.1088/0004-637X/772/1/73}


\bibitem[{{Kumar} \&
  {Panaitescu}(2003)}]{Kumar_Panaitescu_2003_MNRAS.346..905K}
{Kumar}, P., \& {Panaitescu}, A. 2003, \mnras, 346, 905,
  \dodoi{10.1111/j.1365-2966.2003.07138.x}


\bibitem[Levesque et al.(2010)]{2010AJ....139..694L}
Levesque, E.~M., Berger, E., Kewley, L.~J., et al.\ 2010,
\aj, 139, 694. \dodoi{10.1088/0004-6256/139/2/694}


\bibitem[{{Liang} {et~al.}(2004){Liang}, {Dai}, \&
  {Wu}}]{Liang_Dai_2004ApJ...606L..29L}
{Liang}, E.~W., {Dai}, Z.~G., \& {Wu}, X.~F. 2004, \apjl, 606, L29,
  \dodoi{10.1086/421047}

\bibitem[Liang et al.(2006)]{2006ApJ...646..351L} Liang, E.~W., Zhang, B., O'Brien, P.~T., et al.\ 2006, \apj, 646, 351.
    \dodoi{10.1086/504684}

\bibitem[{{Liang} {et~al.}(2015){Liang}, {Lin}, {L{\"u}}, {Lu}, {Zhang}, \&
  {Zhang}}]{Liang_Lin_2015ApJ...813..116L}
{Liang}, E.-W., {Lin}, T.-T., {L{\"u}}, J., {et~al.} 2015, \apj, 813, 116,
  \dodoi{10.1088/0004-637X/813/2/116}

\bibitem[{{Liang} {et~al.}(2010){Liang}, {Yi}, {Zhang}, {L{\"u}}, {Zhang}, \&
  {Zhang}}]{Liang_Yi_2010ApJ...725.2209L}
{Liang}, E.-W., {Yi}, S.-X., {Zhang}, J., {et~al.} 2010, \apj, 725, 2209,
  \dodoi{10.1088/0004-637X/725/2/2209}

\bibitem[{{Liang} {et~al.}(2013){Liang}, {Li}, {Gao}, {Zhang}, {Liang}, {Wu},
  {Yi}, {Dai}, {Tang}, {Chen}, {L{\"u}}, {Zhang}, {Lu}, {L{\"u}}, \&
  {Wei}}]{Liang_Li_2013ApJ...774...13L}
{Liang}, E.-W., {Li}, L., {Gao}, H., {et~al.} 2013, \apj, 774, 13,
  \dodoi{10.1088/0004-637X/774/1/13}

\bibitem[{{Lithwick} \& {Sari}(2001)}]{Lithwick-2001-Sari-ApJ...555..540L}
{Lithwick}, Y., \& {Sari}, R. 2001, \apj, 555, 540, \dodoi{10.1086/321455}

\bibitem[{{L{\"u}} {et~al.}(2012){L{\"u}}, {Zou}, {Lei}, {Zhang}, {Wu}, {Wang},
  {Liang}, \& {L{\"u}}}]{Lu_Zou_2012ApJ...751...49L}
{L{\"u}}, J., {Zou}, Y.-C., {Lei}, W.-H., {et~al.} 2012, \apj, 751, 49,
  \dodoi{10.1088/0004-637X/751/1/49}

\bibitem[{{MacLachlan} {et~al.}(2013){MacLachlan}, {Shenoy}, {Sonbas}, {Dhuga},
  {Cobb}, {Ukwatta}, {Morris}, {Eskandarian}, {Maximon}, \&
  {Parke}}]{MacLachlan-2013-Shenoy-MNRAS.432..857M}
{MacLachlan}, G.~A., {Shenoy}, A., {Sonbas}, E., {et~al.} 2013, \mnras, 432,
  857, \dodoi{10.1093/mnras/stt241}

\bibitem[{{Malacaria} {et~al.}(2021){Malacaria}, {Fletcher}, {Meegan}, \&
  {Fermi GBM Team}}]{Malacaria_Fletcher_2021GCN.29246....1M}
{Malacaria}, C., {Fletcher}, C., {Meegan}, C., \& {Fermi GBM Team}. 2021, GRB
  Coordinates Network, 29246, 1

\bibitem[{{Mangano} {et~al.}(2009){Mangano}, {La Parola}, \&
  {Sbarufatti}}]{Mangano_La_Parola_2009GCN..8767....1M}
{Mangano}, V., {La Parola}, V., \& {Sbarufatti}, B. 2009, GRB Coordinates
  Network, 8767, 1

\bibitem[{{Massey} \& {Johnson}(1998)}]{Massey_Johnson_1998ApJ...505..793M}
{Massey}, P., \& {Johnson}, O. 1998, \apj, 505, 793, \dodoi{10.1086/306199}

\bibitem[{{Melandri} {et~al.}(2010){Melandri}, {Kobayashi}, {Mundell},
  {Guidorzi}, {de Ugarte Postigo}, {Pooley}, {Yoshida}, {Bersier},
  {Castro-Tirado}, {Jel{\'\i}nek}, {Gomboc}, {Gorosabel}, {Kub{\'a}nek},
  {Bremer}, {Winters}, {Steele}, {de Gregorio-Monsalvo}, {Smith},
  {Garc{\'\i}a-Appadoo}, {Sota}, \&
  {Lundgren}}]{Melandri_Kobayashi_2010ApJ...723.1331M}
{Melandri}, A., {Kobayashi}, S., {Mundell}, C.~G., {et~al.} 2010, \apj, 723,
  1331, \dodoi{10.1088/0004-637X/723/2/1331}

\bibitem[{{M{\'e}sz{\'a}ros} \&
  {Rees}(1999)}]{Meszaros_Rees_1999MNRAS.306L..39M}
{M{\'e}sz{\'a}ros}, P., \& {Rees}, M.~J. 1999, \mnras, 306, L39,
  \dodoi{10.1046/j.1365-8711.1999.02800.x}

\bibitem[{{M{\'e}sz{\'a}ros} {et~al.}(1998){M{\'e}sz{\'a}ros}, {Rees}, \&
  {Wijers}}]{Meszaros_Rees_1998ApJ...499..301M}
{M{\'e}sz{\'a}ros}, P., {Rees}, M.~J., \& {Wijers}, R.~A.~M.~J. 1998, \apj,
  499, 301, \dodoi{10.1086/305635}

\bibitem[{{Mimica} {et~al.}(2009){Mimica}, {Giannios}, \&
  {Aloy}}]{Mimica_Giannios_2009A&A...494..879M}
{Mimica}, P., {Giannios}, D., \& {Aloy}, M.~A. 2009, \aap, 494, 879,
  \dodoi{10.1051/0004-6361:200810756}

\bibitem[{{Mizuno} {et~al.}(2009){Mizuno}, {Zhang}, {Giacomazzo}, {Nishikawa},
  {Hardee}, {Nagataki}, \& {Hartmann}}]{Mizuno_Zhang_2009ApJ...690L..47M}
{Mizuno}, Y., {Zhang}, B., {Giacomazzo}, B., {et~al.} 2009, \apjl, 690, L47,
  \dodoi{10.1088/0004-637X/690/1/L47}

\bibitem[{{Mundell} {et~al.}(2007){Mundell}, {Steele}, {Smith}, {Kobayashi},
  {Melandri}, {Guidorzi}, {Gomboc}, {Mottram}, {Clarke}, {Monfardini},
  {Carter}, \& {Bersier}}]{Mundell_Steele_2007Sci...315.1822M}
{Mundell}, C.~G., {Steele}, I.~A., {Smith}, R.~J., {et~al.} 2007, Science, 315,
  1822, \dodoi{10.1126/science.1138484}

\bibitem[{{Panaitescu} {et~al.}(2001){Panaitescu}, {Kumar}, \&
  {Narayan}}]{Panaitescu_Kumar_2001ApJ...561L.171P}
{Panaitescu}, A., {Kumar}, P., \& {Narayan}, R. 2001, \apjl, 561, L171,
  \dodoi{10.1086/324678}

\bibitem[{{Panaitescu} {et~al.}(1998){Panaitescu}, {M{\'e}sz{\'a}ros}, \&
  {Rees}}]{Panaitescu_Meszaros_1998ApJ...503..314P}
{Panaitescu}, A., {M{\'e}sz{\'a}ros}, P., \& {Rees}, M.~J. 1998, \apj, 503,
  314, \dodoi{10.1086/305995}

\bibitem[{{Pei}(1992)}]{Pei_1992ApJ...395..130P}
{Pei}, Y.~C. 1992, \apj, 395, 130, \dodoi{10.1086/171637}

\bibitem[{{Perna} \& {Belczynski}(2002)}]{Perna_Belczynski_2002ApJ...570..252P}
{Perna}, R., \& {Belczynski}, K. 2002, \apj, 570, 252, \dodoi{10.1086/339571}

\bibitem[{{Planck Collaboration} {et~al.}(2016){Planck Collaboration}, {Ade},
  {Aghanim}, {Arnaud}, {Ashdown}, {Aumont}, {Baccigalupi}, {Banday},
  {Barreiro}, {Bartlett}, {Bartolo}, {Battaner}, {Battye}, {Benabed},
  {Beno{\^\i}t}, {Benoit-L{\'e}vy}, {Bernard}, {Bersanelli}, {Bielewicz},
  {Bock}, {Bonaldi}, {Bonavera}, {Bond}, {Borrill}, {Bouchet}, {Boulanger},
  {Bucher}, {Burigana}, {Butler}, {Calabrese}, {Cardoso}, {Catalano},
  {Challinor}, {Chamballu}, {Chary}, {Chiang}, {Chluba}, {Christensen},
  {Church}, {Clements}, {Colombi}, {Colombo}, {Combet}, {Coulais}, {Crill},
  {Curto}, {Cuttaia}, {Danese}, {Davies}, {Davis}, {de Bernardis}, {de Rosa},
  {de Zotti}, {Delabrouille}, {D{\'e}sert}, {Di Valentino}, {Dickinson},
  {Diego}, {Dolag}, {Dole}, {Donzelli}, {Dor{\'e}}, {Douspis}, {Ducout},
  {Dunkley}, {Dupac}, {Efstathiou}, {Elsner}, {En{\ss}lin}, {Eriksen},
  {Farhang}, {Fergusson}, {Finelli}, {Forni}, {Frailis}, {Fraisse},
  {Franceschi}, {Frejsel}, {Galeotta}, {Galli}, {Ganga}, {Gauthier}, {Gerbino},
  {Ghosh}, {Giard}, {Giraud-H{\'e}raud}, {Giusarma}, {Gjerl{\o}w},
  {Gonz{\'a}lez-Nuevo}, {G{\'o}rski}, {Gratton}, {Gregorio}, {Gruppuso},
  {Gudmundsson}, {Hamann}, {Hansen}, {Hanson}, {Harrison}, {Helou},
  {Henrot-Versill{\'e}}, {Hern{\'a}ndez-Monteagudo}, {Herranz}, {Hildebrandt},
  {Hivon}, {Hobson}, {Holmes}, {Hornstrup}, {Hovest}, {Huang}, {Huffenberger},
  {Hurier}, {Jaffe}, {Jaffe}, {Jones}, {Juvela}, {Keih{\"a}nen}, {Keskitalo},
  {Kisner}, {Kneissl}, {Knoche}, {Knox}, {Kunz}, {Kurki-Suonio}, {Lagache},
  {L{\"a}hteenm{\"a}ki}, {Lamarre}, {Lasenby}, {Lattanzi}, {Lawrence}, {Leahy},
  {Leonardi}, {Lesgourgues}, {Levrier}, {Lewis}, {Liguori}, {Lilje},
  {Linden-V{\o}rnle}, {L{\'o}pez-Caniego}, {Lubin}, {Mac{\'\i}as-P{\'e}rez},
  {Maggio}, {Maino}, {Mandolesi}, {Mangilli}, {Marchini}, {Maris}, {Martin},
  {Martinelli}, {Mart{\'\i}nez-Gonz{\'a}lez}, {Masi}, {Matarrese}, {McGehee},
  {Meinhold}, {Melchiorri}, {Melin}, {Mendes}, {Mennella}, {Migliaccio},
  {Millea}, {Mitra}, {Miville-Desch{\^e}nes}, {Moneti}, {Montier}, {Morgante},
  {Mortlock}, {Moss}, {Munshi}, {Murphy}, {Naselsky}, {Nati}, {Natoli},
  {Netterfield}, {N{\o}rgaard-Nielsen}, {Noviello}, {Novikov}, {Novikov},
  {Oxborrow}, {Paci}, {Pagano}, {Pajot}, {Paladini}, {Paoletti}, {Partridge},
  {Pasian}, {Patanchon}, {Pearson}, {Perdereau}, {Perotto}, {Perrotta},
  {Pettorino}, {Piacentini}, {Piat}, {Pierpaoli}, {Pietrobon}, {Plaszczynski},
  {Pointecouteau}, {Polenta}, {Popa}, {Pratt}, {Pr{\'e}zeau}, {Prunet},
  {Puget}, {Rachen}, {Reach}, {Rebolo}, {Reinecke}, {Remazeilles}, {Renault},
  {Renzi}, {Ristorcelli}, {Rocha}, {Rosset}, {Rossetti}, {Roudier},
  {Rouill{\'e} d'Orfeuil}, {Rowan-Robinson}, {Rubi{\~n}o-Mart{\'\i}n},
  {Rusholme}, {Said}, {Salvatelli}, {Salvati}, {Sandri}, {Santos},
  {Savelainen}, {Savini}, {Scott}, {Seiffert}, {Serra}, {Shellard}, {Spencer},
  {Spinelli}, {Stolyarov}, {Stompor}, {Sudiwala}, {Sunyaev}, {Sutton},
  {Suur-Uski}, {Sygnet}, {Tauber}, {Terenzi}, {Toffolatti}, {Tomasi},
  {Tristram}, {Trombetti}, {Tucci}, {Tuovinen}, {T{\"u}rler}, {Umana},
  {Valenziano}, {Valiviita}, {Van Tent}, {Vielva}, {Villa}, {Wade}, {Wandelt},
  {Wehus}, {White}, {White}, {Wilkinson}, {Yvon}, {Zacchei}, \&
  {Zonca}}]{Planck_Collaboration-2016-Ade-A&A...594A..13P}
{Planck Collaboration}, {Ade}, P.~A.~R., {Aghanim}, N., {et~al.} 2016, \aap,
  594, A13, \dodoi{10.1051/0004-6361/201525830}

\bibitem[{{Racusin} {et~al.}(2008){Racusin}, {Karpov}, {Sokolowski}, {Granot},
  {Wu}, {Pal'Shin}, {Covino}, {van der Horst}, {Oates}, {Schady}, {Smith},
  {Cummings}, {Starling}, {Piotrowski}, {Zhang}, {Evans}, {Holland}, {Malek},
  {Page}, {Vetere}, {Margutti}, {Guidorzi}, {Kamble}, {Curran}, {Beardmore},
  {Kouveliotou}, {Mankiewicz}, {Melandri}, {O'Brien}, {Page}, {Piran},
  {Tanvir}, {Wrochna}, {Aptekar}, {Barthelmy}, {Bartolini}, {Beskin}, {Bondar},
  {Bremer}, {Campana}, {Castro-Tirado}, {Cucchiara}, {Cwiok}, {D'Avanzo},
  {D'Elia}, {Della Valle}, {de Ugarte Postigo}, {Dominik}, {Falcone}, {Fiore},
  {Fox}, {Frederiks}, {Fruchter}, {Fugazza}, {Garrett}, {Gehrels},
  {Golenetskii}, {Gomboc}, {Gorosabel}, {Greco}, {Guarnieri}, {Immler},
  {Jelinek}, {Kasprowicz}, {La Parola}, {Levan}, {Mangano}, {Mazets},
  {Molinari}, {Moretti}, {Nawrocki}, {Oleynik}, {Osborne}, {Pagani}, {Pandey},
  {Paragi}, {Perri}, {Piccioni}, {Ramirez-Ruiz}, {Roming}, {Steele}, {Strom},
  {Testa}, {Tosti}, {Ulanov}, {Wiersema}, {Wijers}, {Winters}, {Zarnecki},
  {Zerbi}, {M{\'e}sz{\'a}ros}, {Chincarini}, \&
  {Burrows}}]{Racusin_Karpov_2008Natur.455..183R}
{Racusin}, J.~L., {Karpov}, S.~V., {Sokolowski}, M., {et~al.} 2008, \nat, 455,
  183, \dodoi{10.1038/nature07270}

\bibitem[{{Ren} {et~al.}(2020){Ren}, {Lin}, {Zhang}, {Wang}, {Li}, {Wang}, \&
  {Liang}}]{Ren_Lin_2020ApJ...901L..26R}
{Ren}, J., {Lin}, D.-B., {Zhang}, L.-L., {et~al.} 2020, \apjl, 901, L26,
  \dodoi{10.3847/2041-8213/abb672}

\bibitem[{{Resmi} \& {Zhang}(2016)}]{Resmi_Zhang_2016ApJ...825...48R}
{Resmi}, L., \& {Zhang}, B. 2016, \apj, 825, 48,
  \dodoi{10.3847/0004-637X/825/1/48}

\bibitem[{{Sari} \&
  {Piran}(1999{\natexlab{a}})}]{Sari_Piran_1999ApJ...520..641S}
{Sari}, R., \& {Piran}, T. 1999{\natexlab{a}}, \apj, 520, 641,
  \dodoi{10.1086/307508}

\bibitem[{{Sari} \&
  {Piran}(1999{\natexlab{b}})}]{Sari_Piran_1999ApJ...517L.109S}
---. 1999{\natexlab{b}}, \apjl, 517, L109, \dodoi{10.1086/312039}

\bibitem[{{Sari} {et~al.}(1998){Sari}, {Piran}, \&
  {Narayan}}]{Sari_Piran_1998ApJ...497L..17S}
{Sari}, R., {Piran}, T., \& {Narayan}, R. 1998, \apjl, 497, L17,
  \dodoi{10.1086/311269}

\bibitem[{{Schlafly} \&
  {Finkbeiner}(2011)}]{Schlafly_Finkbeiner_2011ApJ...737..103S}
{Schlafly}, E., \& {Finkbeiner}, D.~P. 2011, \apj, 737, 103
  \dodoi{10.1088/0004-637X/737/2/103}

\bibitem[{{Schlegel} {et~al.}(1998){Schlegel}, {Finkbeiner}, \&
  {Davis}}]{Schlegel_Finkbeiner_1998ApJ...500..525S}
{Schlegel}, D.~J., {Finkbeiner}, D.~P., \& {Davis}, M. 1998, \apj, 500, 525,
  \dodoi{10.1086/305772}


\bibitem[{{Tody}(1986)}]{Tody_1986SPIE..627..733T}
{Tody}, D. 1986, in Society of Photo-Optical Instrumentation Engineers (SPIE)
  Conference Series, Vol. 627, Instrumentation in astronomy VI, ed. D.~L.
  {Crawford}, 733, \dodoi{10.1117/12.968154}

\bibitem[{{Tody} \& {Davis}(1992)}]{Tody_1992_Davis_ASPC...25..484T}
{Tody}, D., \& {Davis}, L.~E. 1992, in Astronomical Society of the Pacific
  Conference Series, Vol.~25, Astronomical Data Analysis Software and Systems
  I, ed. D.~M. {Worrall}, C.~{Biemesderfer}, \& J.~{Barnes}, 484

\bibitem[{{Troja} {et~al.}(2021){Troja}, {Bernardini}, {Breeveld}, {D'Avanzo},
  {Evans}, {Gronwall}, {Gropp}, {Lien}, {Melandri}, {Page}, {Sbarrato}, \&
  {Neil Gehrels Swift Observatory Team}}]{Troja_Bernardini_2021GCN.29233....1T}
{Troja}, E., {Bernardini}, M.~G., {Breeveld}, A.~A., {et~al.} 2021, GRB
  Coordinates Network, 29233, 1

\bibitem[{{Ukwatta} {et~al.}(2010){Ukwatta}, {Stamatikos}, {Dhuga}, {Sakamoto},
  {Barthelmy}, {Eskandarian}, {Gehrels}, {Maximon}, {Norris}, \&
  {Parke}}]{Ukwatta-2010-Stamatikos-ApJ...711.1073U}
{Ukwatta}, T.~N., {Stamatikos}, M., {Dhuga}, K.~S., {et~al.} 2010, \apj, 711,
  1073, \dodoi{10.1088/0004-637X/711/2/1073}

\bibitem[{{Vestrand} {et~al.}(2014){Vestrand}, {Wren}, {Panaitescu}, {Wozniak},
  {Davis}, {Palmer}, {Vianello}, {Omodei}, {Xiong}, {Briggs}, {Elphick},
  {Paciesas}, \& {Rosing}}]{Vestrand_Vestra_nd2014Sci...343...38V}
{Vestrand}, W.~T., {Wren}, J.~A., {Panaitescu}, A., {et~al.} 2014, Science,
  343, 38, \dodoi{10.1126/science.1242316}

\bibitem[{{Wang} {et~al.}(2015){Wang}, {Zhang}, {Liang}, {Gao}, {Li}, {Deng},
  {Qin}, {Tang}, {Kann}, {Ryde}, \& {Kumar}}]{Wang_Zhang_2015ApJS..219....9W}
{Wang}, X.-G., {Zhang}, B., {Liang}, E.-W., {et~al.} 2015, \apjs, 219, 9,
  \dodoi{10.1088/0067-0049/219/1/9}

\bibitem[{{Wei}(2003)}]{Wei-2003-AA...402L...9W}
{Wei}, D.~M. 2003, \aap, 402, L9, \dodoi{10.1051/0004-6361:20030371}

\bibitem[Xin et al.(2011)]{2011MNRAS.410...27X}
Xin, L.-P., Liang, E.-W., Wei, J.-Y., et al.\ 2011,
\mnras, 410, 27. \dodoi{10.1111/j.1365-2966.2010.17419.x}

\bibitem[{{Xin} {et~al.}(2021){Xin}, {Wang}, {Han}, {Wei}, {LI}, {Li}, {Wu},
  {Wang}, {Liang}, {Zhang}, {Qiu}, \& {Deng}}]{Xin_Wang_2021GCN.29235....1X}
{Xin}, L.~P., {Wang}, J., {Han}, X.~H., {et~al.} 2021, GRB Coordinates Network,
  29235, 1

\bibitem[{{Yi} {et~al.}(2013){Yi}, {Wu}, \& {Dai}}]{Yi_Wu_2013ApJ...776..120Y}
{Yi}, S.-X., {Wu}, X.-F., \& {Dai}, Z.-G. 2013, \apj, 776, 120,
  \dodoi{10.1088/0004-637X/776/2/120}

\bibitem[{{Yi} {et~al.}(2020){Yi}, {Wu}, {Zou}, \&
  {Dai}}]{Yi_Wu_2020ApJ...895...94Y}
{Yi}, S.-X., {Wu}, X.-F., {Zou}, Y.-C., \& {Dai}, Z.-G. 2020, \apj, 895, 94,
  \dodoi{10.3847/1538-4357/ab8a53}

  \bibitem[{{Yost} {et~al.}(2003){Yost}, {Harrison}, {Sari}, \&
  {Frail}}]{Yost-2003-Harrison-ApJ...597..459Y}
{Yost}, S.~A., {Harrison}, F.~A., {Sari}, R., \& {Frail}, D.~A. 2003, \apj,
  597, 459, \dodoi{10.1086/378288}

\bibitem[{{Zhang} {et~al.}(2006){Zhang}, {Fan}, {Dyks}, {Kobayashi},
  {M{\'e}sz{\'a}ros}, {Burrows}, {Nousek}, \&
  {Gehrels}}]{Zhang-2006-Fan-ApJ...642..354Z}
{Zhang}, B., {Fan}, Y.~Z., {Dyks}, J., {et~al.} 2006, \apj, 642, 354,
  \dodoi{10.1086/500723}

\bibitem[Zhang et al.(2021)]{2021ApJ...917...95Z}
Zhang, L.-L., Ren, J., Huang, X.-L., et al.\ 2021,
\apj, 917, 95. \dodoi{10.3847/1538-4357/ac0c7f}


\bibitem[{{Zhang} \& {Kobayashi}(2005)}]{Zhang_Kobayashi_2005ApJ...628..315Z}
{Zhang}, B., \& {Kobayashi}, S. 2005, \apj, 628, 315, \dodoi{10.1086/429787}

\bibitem[{{Zhang} {et~al.}(2003){Zhang}, {Kobayashi}, \&
  {M{\'e}sz{\'a}ros}}]{Zhang_Kobayashi_2003ApJ...595..950Z}
{Zhang}, B., {Kobayashi}, S., \& {M{\'e}sz{\'a}ros}, P. 2003, \apj, 595, 950,
  \dodoi{10.1086/377363}

\bibitem[{{Zhang} {et~al.}(2007){Zhang}, {Liang}, {Page}, {Grupe}, {Zhang},
  {Barthelmy}, {Burrows}, {Campana}, {Chincarini}, {Gehrels}, {Kobayashi},
  {M{\'e}sz{\'a}ros}, {Moretti}, {Nousek}, {O'Brien}, {Osborne}, {Roming},
  {Sakamoto}, {Schady}, \& {Willingale}}]{Zhang_Liang_2007ApJ...655..989Z}
{Zhang}, B., {Liang}, E., {Page}, K.~L., {et~al.} 2007, \apj, 655, 989,
  \dodoi{10.1086/510110}

  \bibitem[{{Zheng} {et~al.}(2008){Zheng}, {Deng}, {Zhai}, {Xin}, {Qiu}, {Wang},
  {Lu}, {Wei}, \& {Hu}}]{Zheng-2008-Deng-ChJAA...8..693Z}
{Zheng}, W.-K., {Deng}, J.-S., {Zhai}, M., {et~al.} 2008, \cjaa, 8, 693,
  \dodoi{10.1088/1009-9271/8/6/08}

\bibitem[{{Zhu} {et~al.}(2021){Zhu}, {Fu}, {Liu}, {Xu}, {Gao}, \&
  {Liu}}]{Zhu_Fu_2021GCN.29252....1Z}
{Zhu}, Z.~P., {Fu}, S.~Y., {Liu}, X., {et~al.} 2021, GRB Coordinates Network,
  29252, 1


\end{thebibliography}

\newpage

\begin{longtable}{cccccc}
\caption{Optical Afterglow Photometry Log of GRB~210104A} \\
\hline\hline
$T-T_0$(mid,sec) &  \;Exposure (sec)~\; & \;\;Mag~~~ & \;\;Merr \;\;& Filter & Telescope\\
\hline
\endfirsthead
\multicolumn{6}{c}{(Table 1. continued)}
\endhead
\hline
\endfoot
67	&   10	 &	14.14	&	0.03	&$R$	&	F60A	\\
79	&	10	 &	14.08	&	0.02	&	$R$	&	F60A	\\
91	&	10	 &	13.93	&	0.03	&	$R$	&	F60A	\\
103	&	10	 &	13.78	&	0.02	&	$R$	&	F60A	\\
116	&	10	 &	13.89	&	0.02	&	$R$	&	F60A	\\
128	&	10	 &	13.92	&	0.03	&	$R$	&	F60A	\\
140	&	10	 &	14.01	&	0.03	&	$R$	&	F60A	\\
152	&	10	 &	14.00	&	0.03	&	$R$	&	F60A	\\
656	&	50	&	16.02	&	0.03	&	$R$	&	F60A	\\
708	&	50	&	16.14	&	0.05	&	$R$	&	F60A	\\
760	&	50	&	16.25	&	0.04	&	$R$	&	F60A	\\
813	&	50	&	16.22	&	0.04	&$R$	&	F60A	\\
865	&	50	&	16.28	&	0.04	&	$R$	&	F60A	\\
917	&	50	&	16.35	&	0.05	&	$R$	&	F60A	\\
969	&	50	&	16.50	&	0.05	&	$R$	&	F60A	\\
1021	&	50	&	16.46	&	0.05	&	$R$	&	F60A	\\
1073	&	50	&	16.48	&	0.06	&	$R$	&	F60A	\\
1125	&	50	&	16.46	&	0.08	&	$R$	&	F60A	\\
1950	&	100	&	17.04	&	0.04	&	$R$	&	F60A	\\
2053	&	100	&	17.07	&	0.04	&	$R$	&	F60A	\\
2155	&	100	&	17.12	&	0.05	&	$R$	&	F60A	\\
2257	&	100	&	17.09	&	0.04	&	$R$	&	F60A	\\
2359	&	100	&	17.27	&	0.05	&	$R$	&	F60A	\\
2461	&	100	&	17.23	&	0.05	&	$R$	&	F60A	\\
2681	&	100	&	17.27	&	0.05	&	$R$	&	F60A	\\
2783	&	100	&	17.35	&	0.06	&	$R$	&	F60A	\\
2885	&	100	&	17.29	&	0.06	&	$R$	&	F60A	\\
3127	&	100	&	17.35	&	0.05	&	$R$	&	F60A	\\
3229	&	100	&	17.47	&	0.06	&	$R$	&	F60A	\\
3331	&	100	&	17.46	&	0.06	&	$R$	&	F60A	\\
3433	&	100	&	17.50	&	0.05	&$R$	&	F60A	\\
3638	&	100	&	17.36	&	0.05	&	$R$	&	F60A	\\
3740	&	100	&	17.44	&	0.06	&	$R$	&	F60A	\\
6047	&	150	&	18.01	&	0.09	&	$R$	&	F60A	\\
6199	&	150	&	17.88	&	0.07	&	$R$	&	F60A	\\
6351	&	150	&	17.84	&	0.07	&	$R$	&	F60A	\\
6504	&	150	&	17.90	&	0.08	&	$R$	&	F60A	\\
6656	&	150	&	17.95	&	0.08	& $R$	&	F60A	\\
6938	&	150	&	18.16	&	0.09	&	$R$	&	F60A	\\
7091	&	150	&	18.16	&	0.10	&	$R$	&	F60A	\\
7243	&	150	&	18.01	&	0.08	&	$R$	&	F60A	\\
7395	&	150	&	18.20	&	0.10	&	$R$	&	F60A	\\
7547	&	150	&	18.02	&	0.09	& $R$	&	F60A	\\
7699	&	150	&	18.02	&	0.09	&	$R$	&	F60A	\\
7851	&	150	&	18.11	&	0.09	&	$R$	&	F60A	\\
8003	&	150	&	18.15	&	0.10	&	$R$	&	F60A	\\
8156	&	150	&	17.99	&	0.08	&	$R$	&	F60A	\\
8308	&	150	&	18.26	&	0.10	&	$R$	&	F60A	\\
9316	&	150	&	18.30	&	0.10	&	$R$	&	F60A	\\
9468	&	150	&	18.14	&	0.10	&	$R$	&	F60A	\\
9620	&	150	&	18.22	&	0.10	&	$R$	&	F60A	\\
9772	&	150	&	18.41	&	0.13	&	$R$	&	F60A	\\
9924	&	150	&	18.28	&	0.10	&	$R$	&	F60A	\\
11240	&	150	&	18.55	&	0.16	&	$R$	&	F60A	\\
11392	&	150	&	18.55	&	0.14	&	$R$	&	F60A	\\
11545	&	150	&	18.63	&	0.17	&	$R$	&	F60A	\\
11697	&	150	&	18.70	&	0.18	&	$R$	&	F60A	\\
11849	&	150	&	18.87	&	0.24	&	$R$	&	F60A	\\
12070	&	250	&	18.68	&	0.16	&	$R$	&	F60A	\\
12322	&	250	&	18.39	&	0.12	&	$R$	&	F60A	\\
12574	&	250	&	18.46	&	0.13	&	$R$	&	F60A	\\
12849	&	250	&	18.38	&	0.11	&	$R$	&	F60A	\\
13102	&	250	&	18.88	&	0.17	&	$R$	&	F60A	\\
13354	&	250	&	18.70	&	0.15	&	$R$	&	F60A	\\
17026	&	300	&	19.14	&	0.20	&	$R$	&	F60A	\\
17328	&	300	&	18.71	&	0.14	&	$R$	&	F60A	\\
17630	&	300	&	18.86	&	0.17	&	$R$	&	F60A	\\
18235	&	300	&	19.10	&	0.24	&	$R$	&	F60A	\\
12451	&	200	&	18.56	&	0.10	&	$R$	&	TNT	\\
14037	&	300	&	18.82	&	0.10	&	$R$	&	TNT	\\
14353	&	300	&	18.81	&	0.11	&$R$	&	TNT	\\
16588	&	300	&	19.00	&	0.11	&	$R$	&	TNT	\\
16905	&	300	&	18.94	&	0.11	&$R$	&	TNT	\\
19600	&	1200	&	18.96	&	0.07	& $R$	&	TNT	\\
24704	&	1200	&	19.04	&	0.08	&	$R$	&	TNT	\\
\hline
237	&	30	&	14.15	&	0.02	& $I$	&	F60A	\\
269	&	30	&	14.39	&	0.03	&	$I$	&	F60A	\\
301	&	30	&	14.56	&	0.09	&	$I$	&	F60A	\\
334	&	30	&	14.69	&	0.04	&	$I$	&	F60A	\\
366	&	30	&	14.91	&	0.09	&	$I$	&	F60A	\\
398	&	30	&	14.94	&	0.04	&	$I$I	&	F60A	\\
430	&	30	&	15.04	&	0.06	&	$I$I	&	F60A	\\
462	&	30	&	15.20	&	0.05	&	$I$	&	F60A	\\
494	&	30	&	15.26	&	0.07	&	$I$	&	F60A	\\
526	&	30	&	15.42	&	0.10	&	$I$	&	F60A	\\
1238	&	60	&	16.25	&	0.10	&	$I$	&	F60A	\\
1300	&	60	&	16.22	&	0.08	&	$I$	&	F60A	\\
1362	&	60	&	16.23	&	0.07	&	$I$	&	F60A	\\
1424	&	60	&	16.32	&	0.09	&	$I$	&	F60A	\\
1486	&	60	&	16.42	&	0.10	&	$I$	&	F60A	\\
1549	&	60	&	16.45	&	0.10	&	$I$	&	F60A	\\
1611	&	60	&	16.25	&	0.07	&	$I$	&	F60A	\\
1673	&	60	&	16.42	&	0.09	&	$I$	&	F60A	\\
1735	&	60	&	16.56	&	0.09	&	$I$	&	F60A	\\
1797	&	60	&	16.39	&	0.08	&	$I$	&	F60A	\\
3844	&	100	&	16.95	&	0.09	&	$I$	&	F60A	\\
3946	&	100	&	16.86	&	0.08	&	$I$	&	F60A	\\
4048	&	100	&	17.14	&	0.11	&	$I$	&	F60A	\\
4150	&	100	&	17.13	&	0.11	&	$I$	&	F60A	\\
4355	&	100	&	17.16	&	0.10	&	$I$	&	F60A	\\
4457	&	100	&	17.20	&	0.11	&	$I$	&	F60A	\\
4559	&	100	&	17.20	&	0.10	&$I$	&	F60A	\\
4661	&	100	&	17.53	&	0.17	&	$I$	&	F60A	\\
4865	&	100	&	17.02	&	0.08	&	$I$	&	F60A	\\
5587	&	60	&	17.20	&	0.15	&	$I$	&	F60A	\\
5649	&	60	&	17.39	&	0.17	&	$I$	&	F60A	\\
5711	&	60	&	17.39	&	0.18	&	$I$	&	F60A	\\
5773	&	60	&	17.31	&	0.17	&	$I$	&	F60A	\\
5835	&	60	&	17.00	&	0.13	&	$I$	&	F60A	\\
8966	&	150	&	17.83	&	0.16	&	$I$	&	F60A	\\
9118	&	150	&	17.73	&	0.12	&	$I$	&	F60A	\\
10798	&	200	&	17.94	&	0.15	&	$I$	&	F60A	\\
11001	&	200	&	17.82	&	0.14	&	$I$	&	F60A	\\
13332	&	200	&	18.38	&	0.13	&$I$	&	TNT	\\
13549	&	200	&	18.48	&	0.13	&	$I$	&	TNT	\\
13766	&	200	&	18.28	&	0.12	&	$I$	&	TNT	\\
15633	&	300	&	18.39	&	0.10	&$I$	&	TNT	\\
15950	&	300	&	18.37	&	0.11	&	$I$	&	TNT	\\
16267	&	300	&	18.59	&	0.13	&	$I$	&	TNT	\\
\hline
5305	&	500	&	18.72	&	0.09	&	$B$	&	F60A	\\
9047	&	1000	&	19.31	&	0.08	&	$B$	&	F60A	\\
11246	&	200	&	19.91	&	0.17	&	$B$	&	TNT	\\
11463	&	200	&	19.83	&	0.15	&	$B$	&	TNT	\\
11794	&	200	&	19.73	&	0.17	&	$B$	&	TNT	\\
12011	&	200	&	20.03	&	0.22	&	$B$	&	TNT	\\
12529	&	800	&	19.83	&	0.08	&	$B$	&	TNT	\\
14975	&	900	&	20.04	&	0.10	&$B$	&	TNT	\\
17527	&	900	&	20.16	&	0.13	&	$B$	&	TNT	\\
20079	&	900	&	20.45	&	0.20	&	$B$	&	TNT	\\
23230	&	1800	&	20.64	&	0.16	&	$B$	&	TNT	\\
28932	&	2100	&	20.80	&	0.23	&	$B$	&	TNT	\\
\hline
2678	&	...	&	17.20	&	0.04	& $r$	&	NEXT	\\
2902	&	...	&	17.35	&	0.03	& $r$	&	NEXT	\\
3128	&	...	&	17.36	&	0.04	& $r$	&	NEXT	\\
3351	&	...	&	17.45	&	0.04	&	 $r$	&	NEXT	\\
3572	&	...	&	17.49	&	0.04	& $r$	&	NEXT	\\
3794	&	...	&	17.59	&	0.03	&	 $r$	&	NEXT	\\
4019	&	...	&	17.62	&	0.04	& $r$	&	NEXT	\\
4243	&	...	&	17.54	&	0.05	&	 $r$	&	NEXT	\\
4465	&	...	&	17.61	&	0.04	& $r$	&	NEXT	\\
4692	&	...	&	17.72	&	0.04	&	 $r$	&	NEXT	\\
4923	&	...	&	17.76	&	0.03	& $r$	&	NEXT	\\
5146	&	...	&	17.73	&	0.04	&	 $r$	&	NEXT	\\
5367	&	...	&	17.63	&	0.04	&	 $r$	&	NEXT	\\
5591	&	...	&	17.94	&	0.04	&	 $r$	&	NEXT	\\
5815	&	...	&	17.81	&	0.04	&	 $r$	&	NEXT	\\
6039	&	...	&	17.87	&	0.05	&	 $r$	&	NEXT	\\
6261	&	...	&	17.81	&	0.05	& $r$	&	NEXT	\\
6484	&	...	&	17.87	&	0.04	& $r$	&	NEXT	\\
6707	&	...	&	17.96	&	0.04	& $r$	&	NEXT	\\
6929	&	...	&	17.95	&	0.05	&	 $r$	&	NEXT	\\
7272	&	...	&	18.17	&	0.04	&	 $r$	&	NEXT	\\
7594	&	...	&	17.96	&	0.05	&	 $r$	&	NEXT	\\
7918	&	...	&	18.06	&	0.04	&	 $r$	&	NEXT	\\
8241	&	...	&	18.15	&	0.05	&	 $r$	&	NEXT	\\
8563	&	...	&	18.16	&	0.05	&	 $r$	&	NEXT	\\
8885	&	...	&	18.22	&	0.05	&	 $r$	&	NEXT	\\
9209	&	...	&	18.30	&	0.05	&	 $r$	&	NEXT	\\
9532	&	...	&	18.24	&	0.05	&	 $r$	&	NEXT	\\
9854	&	...	&	18.27	&	0.05	&	 $r$	&	NEXT	\\
10843	&	...	&	18.30	&	0.05	&	 $r$	&	NEXT	\\
11170	&	...	&	18.44	&	0.05	&	 $r$	&	NEXT	\\
11493	&	...	&	18.51	&	0.05	&	 $r$	&	NEXT	\\
11816	&	...	&	18.58	&	0.05	&	 $r$	&	NEXT	\\
12140	&	...	&	18.61	&	0.05	& $r$	&	NEXT	\\
12463	&	...	&	18.44	&	0.05	&	 $r$	&	NEXT	\\
12786	&	...	&	18.42	&	0.05	&	 $r$	&	NEXT	\\
38520	&	...	&	19.52	&	0.07	&	 $r$	&	NEXT	
\label{Tab:opt-data}
\end{longtable}

\clearpage

\begin{deluxetable}{lcccccc|ccc}
\label{tab3}
\tablecaption{Prompt and afterglow emission properties of GRB 210104A in comparison with others GRBs with RS emission detection available in \cite{Japelj_Kopavc_2014ApJ...785...84J} }
\tablehead{
\colhead{GRBs} & \colhead{$z$}  &   \colhead{$\Gamma_{0}$} & \colhead{$E_{\rm \gamma,iso}$}   &  \colhead{$E_{\rm k,iso}$} & \colhead{$R_{B}$}  &  \colhead{$\eta_{\gamma}$}  &  \colhead{$E_{\rm p}$}   &   \colhead{$L_{\gamma,\rm iso}$}   &  \colhead{References}\tablenotemark{a}\\
&   &   &($10^{52}~\rm erg$) &  ($10^{52}~\rm erg$) &  &   &   (keV) &  ($10^{52}~{\rm erg}~{\rm s}^{-1}$)   &}
\startdata
210104A & $0.46$  &  $35$ & $1.30^{+0.20}_{-0.23}$   & $0.39$  &  $28$ & $0.77$ &  $199^{+34}_{-34}$  &  $0.25_{-0.20}^{+0.11}$ &  $This~work$ \\
\hline
990123 & $1.6$   &  $420$ & $239\pm28$   &  $108$ & $1156$  & $0.69$  &  $720^{+10}_{-10}$  &  $8.40^{+1.21}_{-1.21}$  &  $(1)$ \\
\hline
021211 & $1.006$  &  $154$ & $1.10\pm0.13$   &   $3$  & $128$  & $0.27$   &  $47^{+9}_{-7}$    &  $0.25$  &  $(2)$ \\
\hline
061126 & $1.1588$ &  $255$   &   $30\pm3$  &  $12$  & $69$  & $0.71$  & $935^{+360}_{-360}$   &  $2.28$  &  $(3)$  \\
\hline
080319B & $0.9382$  &  $286$  &   $142\pm3$    &   $67.6$  & $16540$  & $0.68$  &  $675^{+22}_{-22}$   &   $10.1^{+0.9}_{-0.9}$  &  $(4)$ \\
\hline
090102 & $1.547$    &   $228$    &   $21.4\pm0.4$   &   $816$  & $6666$  & $0.03$  &  $451^{+73}_{-58}$    &   $7.8$  & $(5)$ \\
\hline
090424 & $0.544$    &     $235$    &   $3.97\pm0.08$  & $258$   &  $25$   & $0.02$  &   $236^{+127}_{-49}$    &  $1.62^{+0.05}_{-0.04}$  &  $(6)$ \\
\hline
130427A & $0.34$    &    $157$     &  $85.0$   &    $521$  &  $4$  &  $0.14$   &  $1028^{+8}_{-8}$   &   $27.0$   &  $(7)$ \\
\hline
\enddata
\tablenotetext{a}{References for $E_{\rm p}$ and $L_{\gamma, \rm iso}$: (1) \cite{Briggs-1999-Band-ApJ...524...82B}, \cite{Guidorzi-2005-Frontera-MNRAS.363..315G};
(2) \cite{Kumar_Panaitescu_2003_MNRAS.346..905K}, \cite{Wei-2003-AA...402L...9W};
(3) \cite{Bellm-2006-Bandstra-GCN..5867....1B}; (4) \cite{Racusin_Karpov_2008Natur.455..183R};
(5) \cite{Golenetskii_2009_Aptekar-GCN..8776....1G}; (6) \cite{Ukwatta-2010-Stamatikos-ApJ...711.1073U}, \cite{Jin-2013-Covino-ApJ...774..114J};
(7) \cite{Huang_Liang_2020ApJ...903L..26H}.}
\end{deluxetable}

\clearpage

\begin{figure}[htbp]
\centering
\includegraphics[angle=0,width=0.9\textwidth]{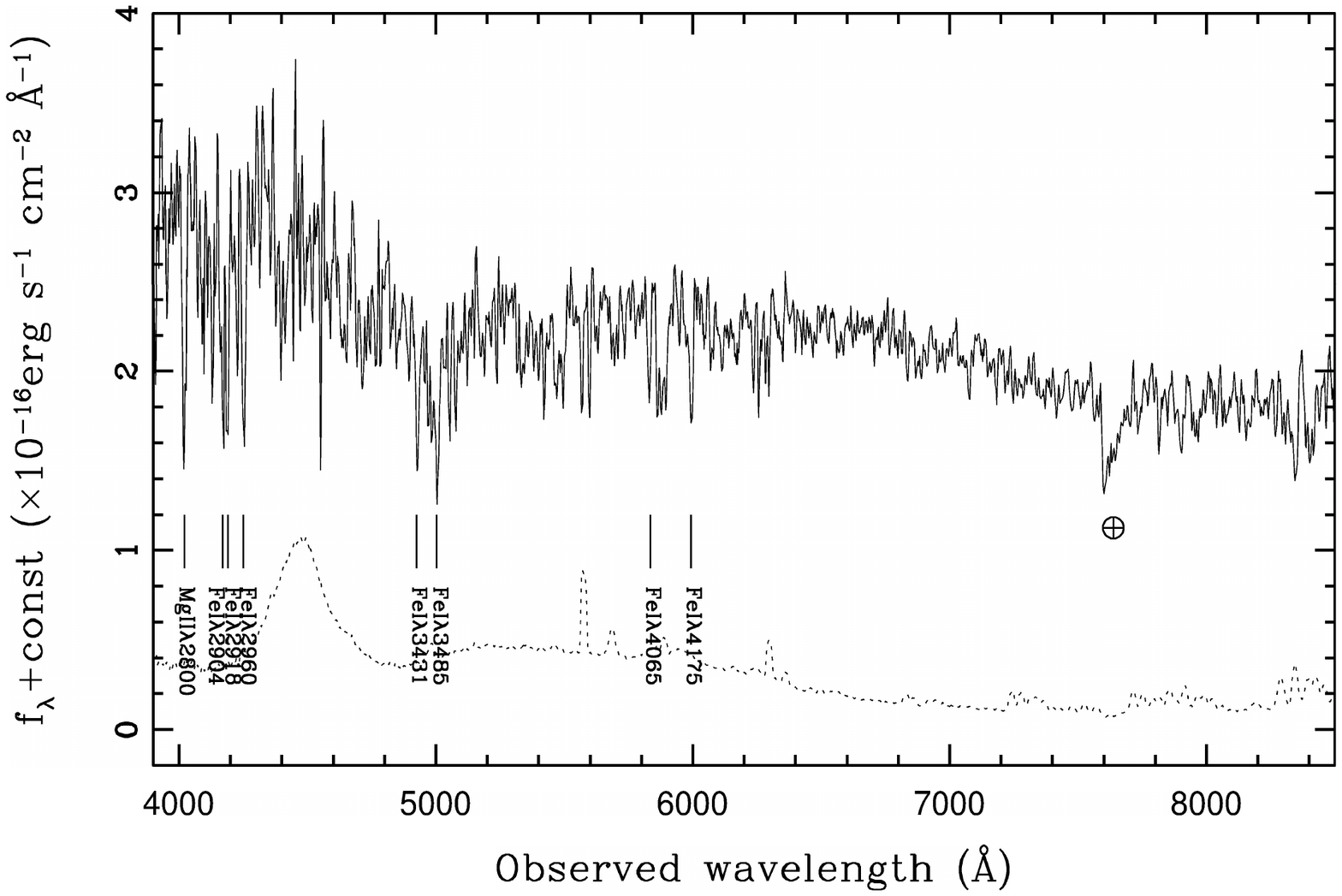}
\caption{Optical spectrum (solid line) of GRB\,210104A in the observer frame obtained with the Xinglong 2.16 m telescope at about 1 hr after the burst trigger. The spectrum is smoothed by a box size of 3\AA. The night sky emission spectrum is shown as dashed line, and the identified absorption features both from the GRB afterglow and the strongest telluric feature at around $\lambda$7600 are also marked. }
\label{Spectrum}
\end{figure}

\begin{figure}[htbp]
 \centering
\includegraphics[angle=0,width=0.45\textwidth]{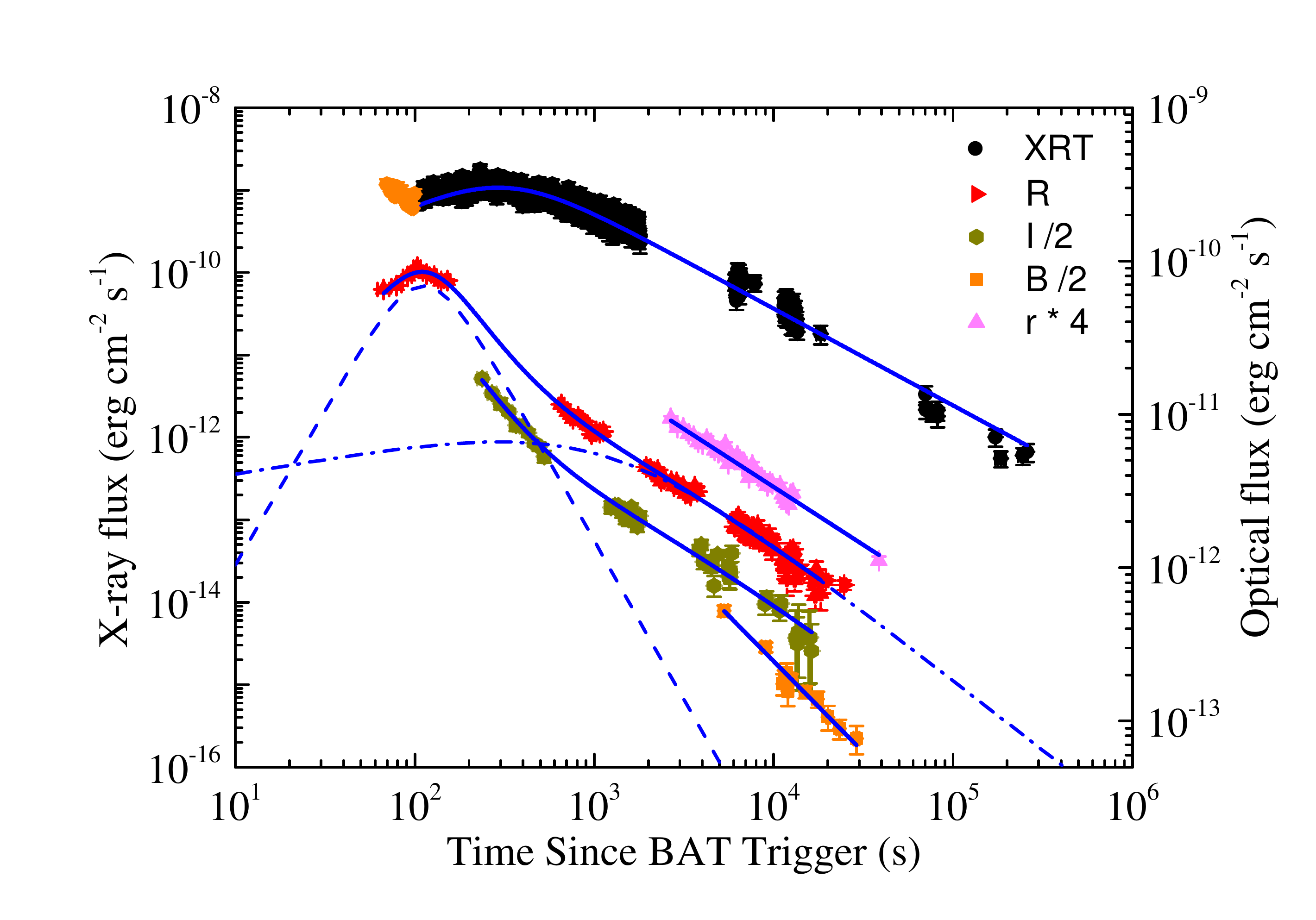}
\includegraphics[angle=0,width=0.45\textwidth]{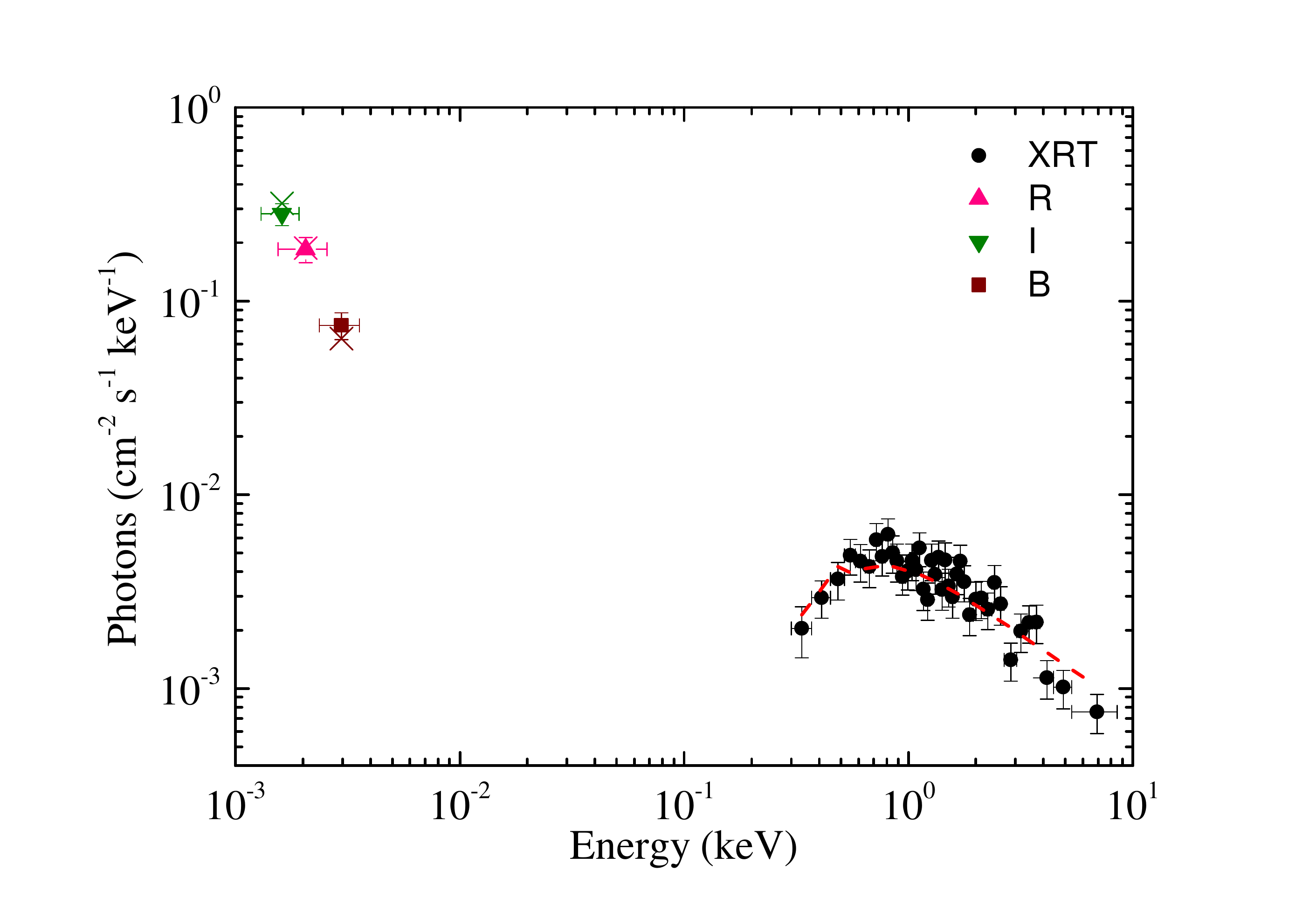}
\caption{
Left panel---Observed optical and X-ray afterglow lightcurves (data dots) of GRB~210104A together with our empirical fits (blue lines).
Right panel---Joint optical-X-ray afterglow observed in the time
interval of $[1.1\sim 1.3]\times 10^4$~s of GRB~210104A
along with our fit with a single power-law function (the red dashed line). The optical data are extinction-corrected for our Galaxy.
Extinction and H I absorption of our Galaxy and the GRB
host galaxy are considered in our spectral fit.}
\label{Photometric}
\end{figure}

\begin{figure}[htbp]
\centering
\includegraphics[angle=0,width=0.9\textwidth]{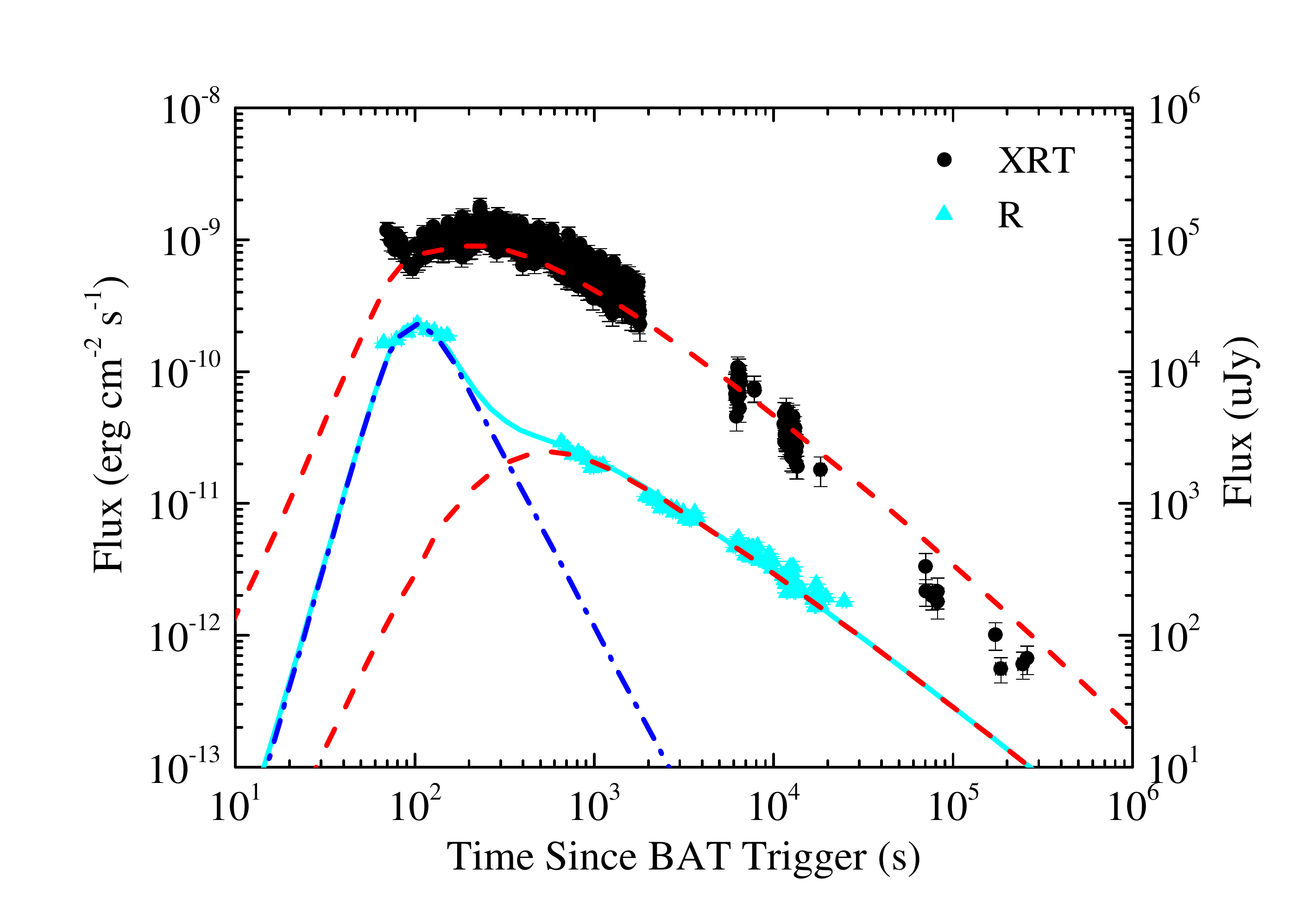}
\caption{Our theoretical fit to the $R$ band and X-ray afterglow lightcurves with the FS and RS models. The RS component is shown with a dashed-dotted line, and the FS component is shown with dashed lines.}
\label{Model_fitting}
\end{figure}

\begin{figure}[htbp]
\centering
\includegraphics[angle=0,width=0.9\textwidth]{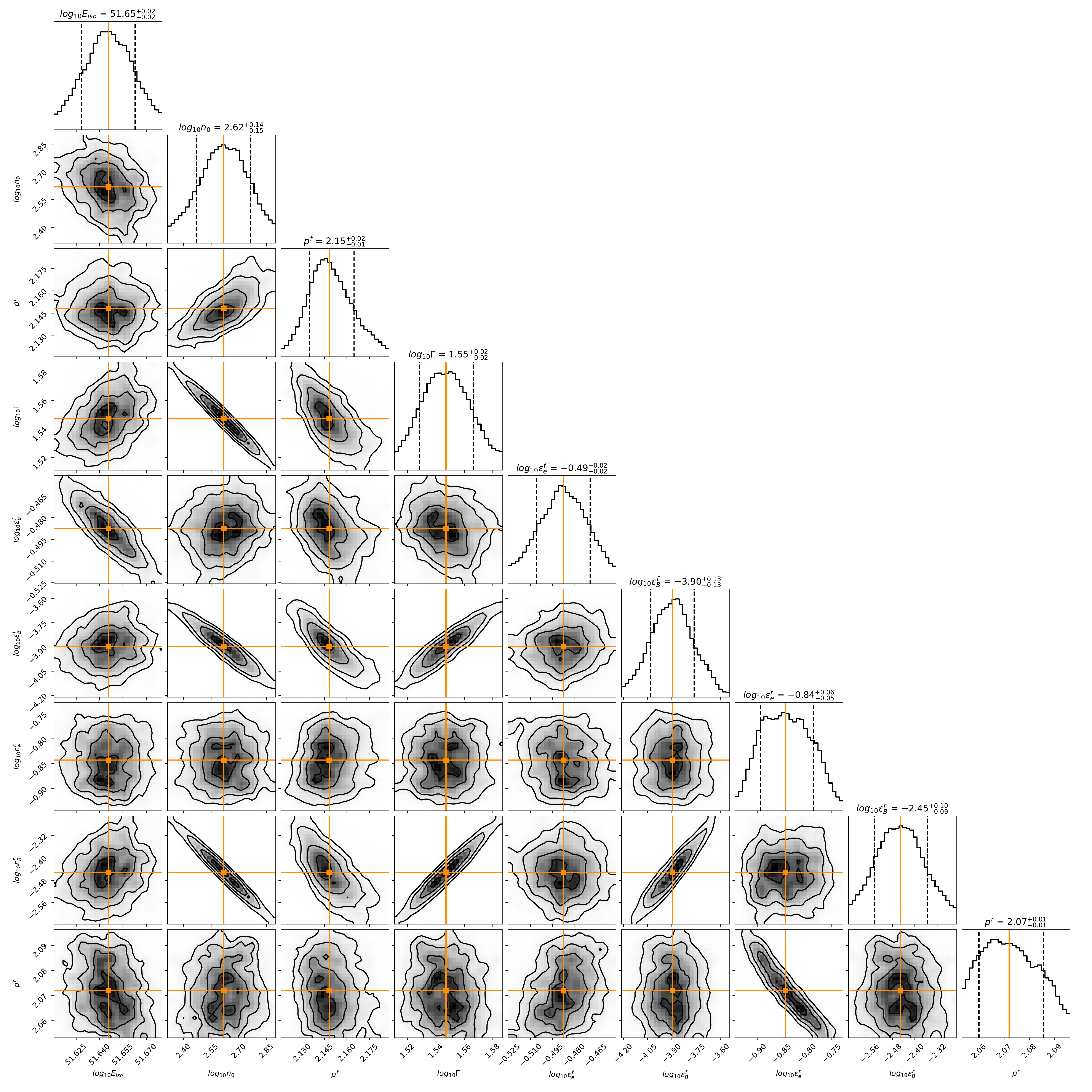}
\caption{Corner plot of the one- and two-dimensional probability distributions of the model parameters derived from our fit to the R band and X-ray afterglow lightcurves of GRB~210104A with the MCMC algorithm. The $1\sigma$ regions centering at the median probability distributions are also reported in the tops and marked with vertical lines in each one-dimensional probability distributions. }
\label{corner}
\end{figure}

\begin{figure}[htbp]
\centering
\includegraphics[angle=0,width=0.9\textwidth]{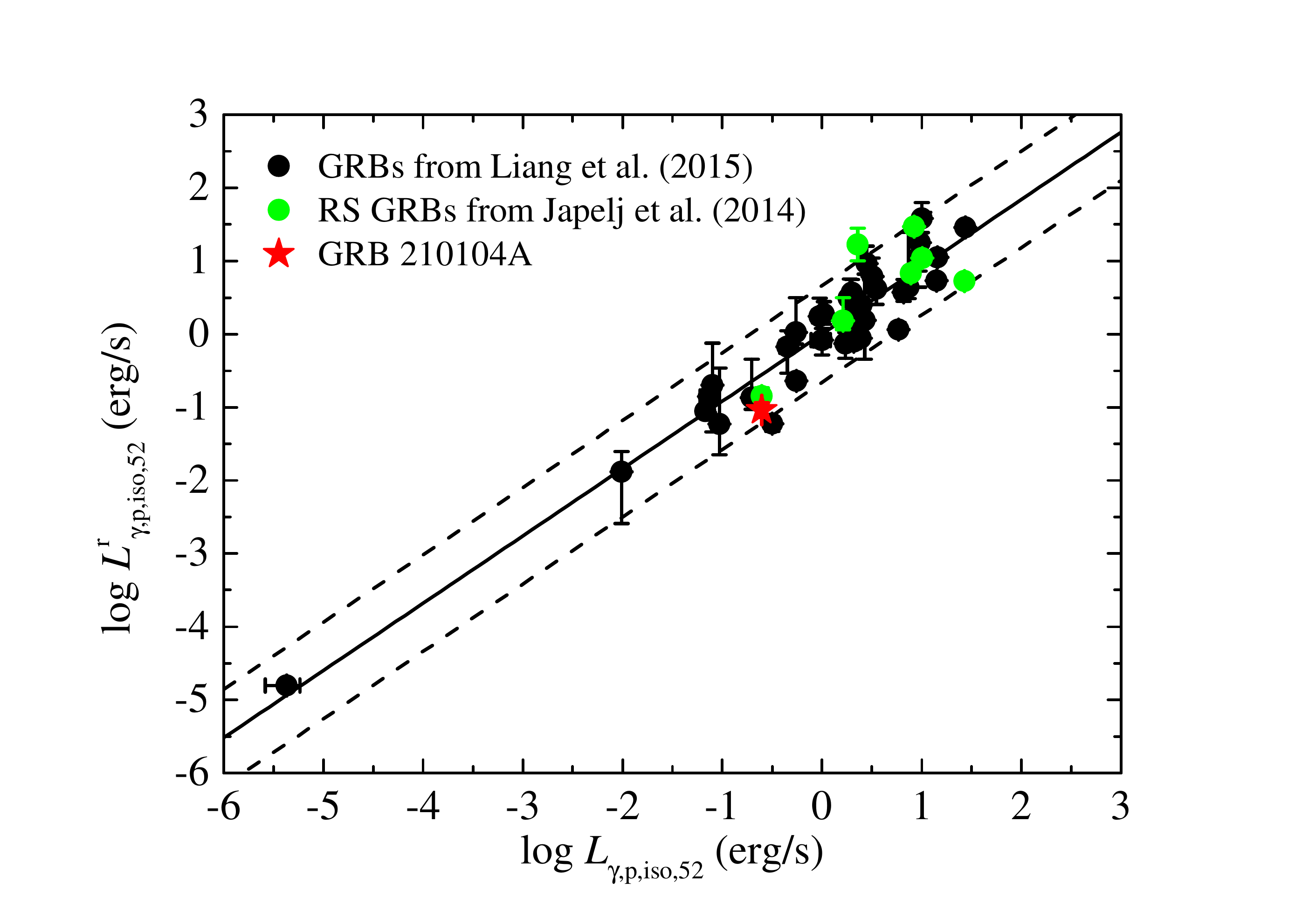}
\caption{Luminosity calculated with the $L_{\gamma, \rm iso}-E_{\rm p,z}-\Gamma_{\rm 0}$ relation reported by \cite{Liang_Lin_2015ApJ...813..116L} as a function of the observed luminosity for GRB 210104A and other GRBs from \cite{Japelj_Kopavc_2014ApJ...785...84J} as marked in the plot. The solid and dashed lines mark the relation and its $2\sigma$ dispersion.}
\label{Correlation}
\end{figure}

\clearpage

\end{document}